\documentclass[a4paper,12pt]{article}
\usepackage{amsmath}
\usepackage{amsfonts}
\usepackage{amssymb}
\usepackage{epsfig}
\usepackage{bbm}
\usepackage{latexsym}
\usepackage{graphicx}
\usepackage{fancybox}
\usepackage{mathrsfs}

\newcommand{\p} \prime
\newcommand{\e} \epsilon
\newcommand{\la} \lambda

\newcommand{\LV}{\Lambda_{LV}}
\newcommand{\sign} {\rm sign}
\newcommand{\llv} {\Lambda_{LV}}
\newcommand{\om} \omega   \newcommand{\Om} \Omega

\newcommand{\al} \alpha
\newcommand{\bt} \beta
\newcommand{\be} {\begin{equation}}
\newcommand{\ee} {\end{equation}}
\newcommand{\ba} {\begin{eqnarray}}
\newcommand{\ea} {\end{eqnarray}}
\def\lrD{\mathrel{{\cal D}\kern-1.em\raise1.75ex\hbox{$\leftrightarrow$}}}
\def\lr #1{\mathrel{#1\kern-1.25em\raise1.75ex\hbox{$\leftrightarrow$}}}

\begin{document}

\vskip 1cm
\begin{center}
   {\large  \sc Constructing QFT's \\ wherein Lorentz Invariance
   is broken\\ by dissipative effects in the UV\footnote{\it This work
   has been presented in the SISSA conference {\rm From  Quantum to Emergent 
   Gravity} 
   in June 2007.
   It contains large extracts of the manuscript 
   {\rm 
   Confronting the trans-Planckian question of inflationary cosmology
    with dissipative effects. }}}
\end{center}

\begin{center}
       R. Parentani 
\end{center}

\vskip 1cm

\begin{center}
{\it Laboratoire de Physique Th\'eorique, UMR 8627},\\ 
Universit\'e  Paris-Sud 11, B\^at. 210,\\ 91405 Orsay Cedex, France
\end{center}

\vskip 1cm
%

\begin{abstract}

\noindent
There has been a recent interest
in considering Quantum Field Theories 
in which Lorentz Invariance is broken in the UV sector.
However 
attention has been mostly limited 
to dispersive theories.
In this work we provide the generalized settings for 
studying dissipation. 
Unitarity is preserved by coupling the original fields
to additional (heavy) 
fields which induce the dissipation. 
Starting with Lagrangians breaking LI in the UV,
we learn that dissipative effects unavoidably develop in the
effective theory.
We then covariantize these Lagrangians 
in order to 
address 
the trans-Planckian question of 
inflation and black hole physics. The peculiar properties of the additional
fields inducing dissipation is revealed by the covariantization.
The links with the phenomenological approach to Quantum Gravity
and with some Brane World scenarios are also discussed. 
\end{abstract}
\vskip .3 truecm


%
%
%
\newpage

\section{Introduction}

 Even though relativistic QFT provides an excellent description of
 particle physics, being non-compact, Lorentz symmetry has been 
  tested only up to a certain energy scale\cite{Jacobson:1991gr}. 
Thus one cannot exclude the possibility that some 
high energy processes 
break the invariance under boosts,
thereby introducing a threshold energy $\LV$,
and a preferred
frame.
It is therefore of interest to understand what would be  
the signatures when 
this possibility is realized.
 
 Unlike dispersive effects which have been studied in 
 details~\cite{Jacobson:2005bg}, 
 dissipative effects have so far received much less attention.
In this work we provide the grounds for this extension. 
To this end, we 
first construct QFT displaying dissipative effects
in the UV. 
Indeed, 
to handle dissipation requires 
settings which are wider than those of 
relativistic and dispersive theories 
because,  
 if one introduces dissipation
from the outset in the usual settings, one looses both 
unitarity and 
predictability.
To preserve them, 
we shall therefore work with Hamiltonian theories in which dissipative effects
are caused by interactions with additional degrees of freedom.
Doing so, we shall discover that 
dissipative effects are {\it generic}. That is, when starting with a
bare Lagrangian in which LI is broken by some kinetical or interaction term 
in the UV, the effective theory (the generating functional) unavoidably
develops dissipation above a certain energy scale, simply because nothing
can prevent this. (With relativistic QFT instead, LI did prevent it).
We also 
learn that it is 
illegitimate to deal with purely dispersive QFT, since these do not satisfy
Kramer's relations.

Being engaged in a  procedure of generalization, we should have a clear
idea of our motivations, aims, and requirements.
We first present 
these 
aspects.

\subsection{Two requirements}

\indent 

{\it $1^{st}$. 
Unitarity}

\noindent
As already mentioned,
we require that our QFT evolve unitarily. 
This 
requirement implies that 
the dissipative effects be 
produced by the Hamiltonian dynamics of the entire system.
In other words, we shall introduce 
additional degrees of freedom 
called 
$\Psi$, which play the role of an environment,
and couple them to the original field $\phi$ in such a way that the latter
develops dissipative effects.
This means that the (dressed) two-point function of $\phi$ will
be given by the usual QM trace
\be
G_W(x,y)= {\rm Tr}
\, \Big[  \hat \rho_T \, \hat \phi(x) \, \hat \phi(y)  \big],
\label{Gw}\ee
where $\hat \rho_T$ is the initial 
matrix density of the entire system
$\Psi + \phi$, where 
$\hat \phi(x)$ is the Heisenberg field operator evolved with the 
time ordered exponential of 
the total Hamiltonian, 
and where the trace is taken over both $\Psi$ and $\phi$.

\vskip .2 truecm

{\it $2^{d}$. Stationarity and homogeneity}

\noindent
Our second requirement concerns the properties of dissipative effects.
When considering the theory in vacuum and in Minkowski space-time, 
we impose that the dissipative effects become significant only above 
a critical 
energy $\LV$, and that they preserve  the stationarity, the homogeneity, and
isotropy of flat space-time. 
In this case, dissipative effects
define a preferred frame 
which is inertial and globally defined.

Then, irrespectively of the properties of the 
additional degrees of freedom $\Psi$
and their interactions with 
$\phi$, 
the Fourier transform of the retarded 
Green function $G_R(x,y)$, which is given by
$G_R(x,y) = \theta(t_x - t_y)\, 2$Im$G_W(x,y)$, 
where $G_W$ is the 
Wightman function of eq. (\ref{Gw}), 
is of the form
\be
G_R(\om, p) = 
{-i \over \Big(-\om^2 + p^2 + \Sigma_R(\om,p)  \Big)} \, .
\label{GfF}
\ee
In 
the true vacuum, at the level of the 2-pt functions,
the dissipative (dispersive) effects 
 are indeed completely characterized by the imaginary and odd (real and even) 
 part in $\om$ 
of the (retarded) self-energy $\Sigma_R(\om,p)$. 
 
 In these expressions, the energy $\om$ and the spatial momentum
 square $p^2$ have been defined in the preferred frame. 
To prepare the covariantization of our theory, and therefore its extension to 
curved space-time, it is usefull to   
characterize the preferred frame 
in a coordinate invariant way by a unit time-like 
vector field, here after called $l^\mu$. 
 Then $\om$ and $p^2$  are given by 
 \be
 \om \equiv   l^\mu \, p_\mu \, , 
 \quad  p^2 \equiv  \perp^{\mu\nu} \, p_\mu p_\nu \ ,
 \label{perp}
 \ee
 where $\perp^{\mu\nu} \equiv \eta^{\mu\nu} + l^\mu l^\nu$ is the 
 (positive definite) metric
 in the spatial sections orthogonal to $l^\mu$. 
  
  The novelty 
 is that $\Sigma_R$ is  a function of $\om$ and $p$ separately,
 and not only  of the relativistic invariant $\om^2 - p^2$
 as it is the 
  case in relativistic QFT. 
 When $\Sigma$ depends on both  $\om$ and $p$,
 dissipation can become significant above a critical
 energy {\it on the mass shell}, i.e. along the 
 minima of the denominator of eq. (\ref{GfF}); 
 a possibility forbidden in LI theories.

 With the observation of the simplicity of eq. (\ref{GfF}),
 we understand that most of 
  the properties of $\Psi$ will be irrelevant
 when restricting attention to observables built with only $\phi$.
  In other words,
  the effective action of $\phi$ only retains little information about $\Psi$
 in 
 its moments, 
  the second of which being the self-energy $\Sigma$.
In this paper
we shall exploit 
this freedom 
by choosing the simplest models of $\Psi$ which 
deliver the required properties of 
$\Sigma$.
 
\subsection{Aims and Motivations} 

%
\subsubsection{The phenomenology and the links with Quantum Gravity}

 \noindent
We first aim to describe the phenomenology of dissipative effects 
in Minkowski space-time and in vacuo.
This is rather easy. From eq. (\ref{GfF})
we understand that the {phenomenology} of dissipative effects
 respecting our (minimal) restrictions will be 
 governed by a limited set of functions. In fact, at the level of 2-pt functions,
  the imaginary part of $\Sigma$ is the only relevant quantity. 
 One easily verifies that 
  the following self-energies induce
 significant dissipation only above $\LV$
\be
{\rm Im}\Sigma^{(n)}_R(\om, p) 
    = - {\om \over \llv}
    \, p^2 
   \left({p^2 \over \llv^2}\right)^{n} .
\label{ndiss}
\ee
Later in the text, we shall provide 
Lagrangians of $\Psi$ and $\phi$ 
which produce this class of imaginary self-energies.
We can already relate this class
to the set of
non-linear dispersion relations which have been used 
in phenomenological studies:
 \be
 \omega^2 = F_n^2(p^2) = 
 p^2 \,  \pm  p^2 \left({p^2 \over \llv^2}\right)^{n}
  + O\Big((p^{2})^{n+2}\Big)\, .
 \label{disprel}\ee
The $+$ ($-$) sign gives
  superluminous (subluminous) propagation. 
 It is also worth noticing that a phenomenological study
 of dispersive and dissipative effects taken toghether can be done with 
 eq. (\ref{GfF}) by considering both Re$\Sigma_R$ and Im$\Sigma_R$ non zero.

\vskip .1 truecm


\noindent
Even though Quantum Gravity is not part of our main concerns,
let us say a few words about the relations with our work.

First, one should clearly distinguish two possible scenarios
in which LI might be broken. 
First, the violation
could be {\it fundamental}. In this case, 
the violation will show up in every circumstances, 
including in Minkowski space-time 
(or what replaces it) and in vacuum. Secondly, 
the violation could be only {\it effective}. That is, it could be induced 
by radiative corrections in some circumstances, e.g. in black hole
geometries, in cosmological backgrounds, or in thermal states. This line
of thought have been developped in \cite{Parentani:2007mb,Arteaga:2003we}

Secondly, Quantum Gravity, whatever version is adopted, 
implies that the smooth manifold of GR should be replaced by 
a new structure when reaching a UV threshold energy, call
it $\LV$. When adopting a phenomenological point of view,
there is a shift in the interest. The 
question is no
longer: What is this new structure~? but rather: How would it manifest
itself in observables~? 
That is to say: What are the new expressions of the $n$-point correlation
functions~? Remember that the predictions of a QFT 
are all based on its correlation functions.

The simplest of these are the 2-point functions.
In whatever replaces the Minkowski manifold, 
spatial homogeneity and stationarity will be preserved in the mean,
(otherwise that version of QG would not describe our world).
This implies that the Fourier transform of the 
"true" Feynman Green function\footnote{Whatever QG may be,
this function will be given by an expression {similar} to eq. (\ref{Gw})
with the trace taken over the "true" degrees of freedom.
{\it For the benefit of the reader,
 I add here a remark made by a 
 referee concerning this sentence and my reply to it: 
"While of course one might claim this is a plausible scenario, 
one might as well conjecture that in no way transplanckian physics 
can be described by a quantum field theory of any sort 
(e.g. QM could be emergent as well)." My answer 
is twofold.
First, I wrote "similar" 
precisely to leave open the possibility that 
the trace be not "the usual QM" one. 
Second, even if QM is only emergent, when considering the 2-pt 
function of accessible (effective) degrees of freedom, 
as $\phi$ in the present case,
(or like the 2-pt function of the phonon field in a BEC),
 the true expression will, {at least},
 contain a trace (to average out the configurations so that the
 correlation function only depends on the arguments of the two operators),
 a state function (to weigh these configurations), and two operators (which 
 might be complicated composite objects), 
 because, in the IR, the 2-pt function must behave according to QM.
 The beauty of 2-pt correlation functions is that they are
 c-number functions depending only $\omega$ and $\bf p$, both in the
 effective description and in the true description. Therefore 
 a comparison of two versions directly delivers the relevant modifications,
  see \cite{Parentani:1996ta}
   for a comparison of 2pt functions obtained from 
   QFT  in a fixed background and 
 from solutions of the Wheeler-DeWitt equation.}}
%
%
%
%
will behave as 
in eq. (\ref{GfF}).\footnote{It
has been claimed that 
the reparametrization invariance of QG would 
severly restrict the 
 true observables 
 (e.g. to be topological charges). 
If this could be correct in pure 2+1 QG\cite{Witten:1988hc}, 
it is harder to conceive the relevance 
of this statement when 
applied 
to 
4D, 
both from a phenomenological
and a theoretical point of view. 
At a theoretical level, the 
quantum transitions of (heavy) atoms exchanging photons 
are governed by the 2-pt functions of the radiation
field evaluated where the atoms sit, given their wave function.
This should remain true when including QG effects, 
see \cite{Parentani:1996sz} for a description of atomic transitions
based on solutions of the WDW equation, i.e. when reparametrization 
invariance has been taken into account.} 
 (Of course LI might still be respected in which case the
self-energy induced by QG will only be a function of $\om^2 - p^2$.)

The link between our approach 
and the phenomenological approach to QG
which consists in parametrizing its effects (rather than computing them
from first principles) is clear: Since we provide the  
general expression of the 2pt function compatible with QM, 
in the sense that the Equal Time Commutations relations are 
 satisfied,
our expressions can also be used in a phenomenological QG perspective.

%

%
\subsubsection{Mode creation in expanding universes}

\noindent
When assuming that LI is broken in the UV in Minkowski space, 
one can then also assume that the density of degrees of freedom, 
or modes, is finite and of the order of $\llv^3$.
(This density should not 
be confused with the density of quanta which is always finite in decent quantum states, the "Hadamard" states\cite{Birrell:1982ix,Wald:1995yp}.) 
This option was not available as long as LI prevailed, 
and was in fact the source of 
the UV divergences. 
If the cutoff $\llv$ is much higher than the typical frequencies
involved in the observables we have access to, 
it is rather easy to show that the expectation value of these observables
will not be significantly affected by $\llv$, a simple example which can be 
worked out explicitely is furnished by the Casimir effect. Therefore, 
in Minkowski space, one cannot expect any significant deviation
induced by having cutoff the mode density.

On the contrary, it is more challenging to consider how would such a theory 
behave in an expanding universe. Indeed one faces an alternative.
Either the total number of degrees of freedom in a given comoving volume
would stay fixed (or nearly fixed), and therefore the density would decrease 
like the inverse of the proper volume,
or it is the density that would stay (nearly) fixed, and in this case,
the number of degrees of freedom would grow linearly with the proper volume, thereby implying mode creation. 
 
 The first alternative seems already excluded because the
 volume of our visible universe increased at least by a factor 
 of $e^{360}$ since the onset of inflation
 ($360$ = $60$ e-foldings during inflation $\times$ 
 two for the radiation
 era $\times$ three because of 3D). Indeed either the density was
 absurdly high at the onset of inflation, or if
 the intial density was decent,  we should today be lacking
 degrees of freedom.
 Moreover if it were true, one could measure the growth of the scale 
 factor by probing locally the vacuum, thereby violating the 
 Equivalence Principle.
 So, once having assumed a finite density in flat space,
  we are left with the conclusion that mode creation 
  is unavoidable in expanding universes, and therefore in any curved background
  geometry.
 
 When adopting this second alternative, 
 two questions should be confronted:
  
\noindent
  How to describe mode creation in QFT ?  
   
\noindent
   What fixes the
 state of the newly born modes ?
 
Our motivation is to 
 confront these two questions
 in the presence of 
 dissipative effects. 
  To this end we first need to extend our 
  QFTs to curved space times. 
  
  The main principle we adopt 
 is the Equivalence Principle (or 
more precisely its extension
  in the presence of the vector field $l^\mu$).
 That is, 
the Lagrangians 
will be a sum of scalar functions
of the four local fields $\phi$, $\Psi$, $g_{\mu \nu}$ and $l^\nu$
which reduce locally to their Minkowski value in the zero curvature limit.
 This principle 
 fixes the action density (up to the possibility of some non-minimal 
 coupling) and determines the tensorial nature of the  $\Psi$ fields. 
In the models we shall use, 
the $\Psi$ fields form
a {\it dense} set of local degrees of freedom at rest 
 with 
 $l^\mu$.
 
 When 
 the Equivalence Principle is respected, 
 without fine tuning nor additional hypothesis,
 we shall see that dissipative 
 QFT
are such that, as the universe expands,
 underdamped modes emerge from overdamped modes 
 --thereby effectively describing the creation 
 of propagating degrees of freedom-- 
 in such a way that their (proper) density stays constant. 
 Moreover, unlike in former attempts to describe 
 "mode creation"\cite{Foster:2004yc,Kempf:2000ac}, 
 there is no need to supply 
 an extra condition to fix the state of the newly born modes.
 Indeed, since our models are Hamiltonian,
 all the information is contained in the 
 initial density matrix $\hat \rho_T$. 
 It should also be noticed 
that when considering contracting universes,
 or locally contracting regions, unitary models 
 imply the inverse process, i.e. 
 "mode destruction", whereas it is unclear what 
 the settings of \cite{Foster:2004yc,Kempf:2000ac} predict in this case.
 As a last comment, the mechanism at work in unitary models can also 
 be 
 described
 as "mode conversion" by analogy 
  to processes encountered
 in condensed matter.\footnote{To obtain
 dissipative QFT,  
we could have 
searched for inspiration in quantum optics 
or
other condensed matter models since the effective LI is also
broken above a certain scale.
We have chosen not to pursue this 
approach 
for several reasons.
First 
this approach 
lacks generality 
and thus hides the steps necessary to {construct} dissipative models. 
The second reason comes 
from the Equivalence Principle 
considered in expanding universes. Indeed any QFT emerging from a 
{\it discrete} structure, e.g. a set of atoms, is bound to violate 
the Equivalence Principle after a large number 
of e-folds. We see no way to escape this
conclusion.}
 
 \subsubsection{The trans-Planckian question}

\noindent
In inflationary cosmology, the primordial density 
fluctuations arise from the amplification of vacuum 
fluctuations which had very short wave lengths
(very large proper frequencies) at the onset of inflation\cite{Mukhanov:1990me}.
Similarly, Hawking radiation emitted by a black hole
emerges from configurations which had extremely high 
initial frequencies\cite{PhysRep}. In both cases, the unbounded frequency growth 
questions the validity of the predictions because these 
have been obtained using the standard treatment,
namely QFT 
in a curved space time. 
However there is 
no reason to believe that these settings still provide a reliable
approximation for 
frequencies way above 
the Planck scale. 
 
Following original work of Unruh 
and Jacobson\cite{Unruh:1994je,Jacobson:1993hn}, the 
robustness of the standard predictions against modifying the theory in the UV
have been tested by introducing dispersion relations which 
become non linear above a certain UV scale $\LV$, see
eq. (\ref{disprel}).
Even though the propagation of the configurations is severely 
modified when this scale is reached, 
it was shown that the properties 
are essentially unmodified
when the two relevant scales 
are well separated, i.e. in BH physics\cite{Brout:1995wp} 
 when $\kappa/\LV \ll 1$ where $\kappa$
is the surface gravity, 
and in 
inflation\cite{Martin:2000xs,Niemeyer:2000eh} when $H/\LV \ll 1$ where $H$ is 
the Hubble parameter. What guarantees the robustness 
is that the vacuum state evolves adiabatically. 

Our aim is to generalize these works
by providing dissipative models in which
the power spectrum (and Hawking radiation) can be 
computed. These 
models are the same as those of the former subsection.
The fact that the Equivalence Principle is preserved
will guarantee, as we shall see, 
the adiabaticity of the evolution of the 
true vacuum as long as the gradients of the metric are much smaller
than the UV scale $\LV$. 
  
 To conclude, 
 we would like bring to the reader's attention 
 to the following remark. To 
 obtain the power spectrum in inflationary models,
 and 
 the asymptotic properties of Hawking radiation,
 it is always {\it sufficient} to know the 2pt function of eq. (\ref{Gw})
 since 
 \ba
 {P}_k(t) &\equiv & \int d^3x \, e^{-i {\bf k x}} \,  G_{a}(t,{\bf x}; t, {\bf 0}),
 \nonumber \\
 n_{\om, l, m} + \frac{1}{2} 
 &\equiv& \int dt \, e^{i \om t} \,   G_{a}(t, r; 0, r, l,m), \quad r \gg r_h .
 \label{PandS}\ea
 In inflation, the first equality
 follows from the definition of the power spectrum $P_k$ which 
 is given by the spatial Fourier transform of the anti-commutator $G_a$
 evaluated, at equal time after horizon exit, in the Bunch-Davies vacuum. 
 As of Hawking radiation, $n(\om, l, m)$,
 the asymptotic distribution at fixed angular momentum $l,m$,
  is given by Fourier transform with respect 
 to the asymptotic time of the in-vacuum 
 anti-commutator far away from the hole. 

 We now remark that 
 the knowledge of $G_a$ 
 becomes 
 {\it necessary} in the presence of dissipation because 
  the only way to obtain $P_k$ and $n(\om)$
 is through the above expressions. 

\vskip .2 truecm

We have organized the paper as follows.
We first 
construct dissipative theories 
in Minkowski space-time. In this case, because of stationarity
the analysis is simple 
and instructive. 
We then covariantize these models and briefly
comment on the trans-Planckian question in inflation 
and in black hole physics. In long Appendices 
we provide self-contained presentations of the properties of 
2-pt functions in the presence of dissipation.

\section{Dissipation in Minkowski space from $LV$ effects}

In this section, we provide 
a class of models defined in Minkowski
space time which exhibit dissipative effects above
a certain energy scale $\llv$. Stationarity, homogeneity
and isotropy will be exactly preserved. Therefore, 
the only invariance
of relativistc QFT which is 
broken is that under boosts.
These theories define a preferred rest frame which is globally defined, 
as it is the case of 
FLRW space-times. 
Even though a covariant description 
 exists, for simplicity of the presentation, 
we first work in that frame. At the end of this Section
we shall covariantize them to prepare the extension to curved space-times. 

\subsection{Free field settings}

We start with a brief presentation of the free field description
to introduce our notations and to point out what are the 
properties which are 
lost in the presence of dissipation.

The action of our free massless field $\phi$ is the usual one:
\be
S_\phi = {1 \over 2} \int dt \, d^3{x} \, (\partial_t\phi\, \partial_t\phi -
 \partial_{\bf x}\phi \cdot  \partial_{\bf x}\phi  ) \, ,
 \label{freeS}
\ee 
Due to the homogeneity of space,
the equation of motion can be analyzed mode by mode:
\be
\hat \phi(t,{\bf x}) = \int\! {d^3{p}\over (2 \pi)^{3/2}}\,  e^{i\bf p \cdot x} 
\hat \phi_{\bf p}(t) 
\, .
\ee
The mode operator $\hat \phi_{\bf p}(t)$, 
the Fourier transform of the field operator, obeys
\be
(\partial^2_t + \om_p^2 ) \hat\phi_{\bf p} =0\, ,
\label{eqom}
\ee
where $\om_p^2= p^2 = \bf p \cdot p$ is the standard relativistic dispersion relation.
Notice that 
eq. (\ref{eqom}) is second order, homogeneous (no source term),
and time reversible (no odd power of $\partial_t$), three
properties we shall loose when introducing
interactions breaking $LI$.

In homogeneous space-times, 
the canonical Equal Time Commutator of the field and its momentum
implies 
\be
[\hat\phi_{\bf p}(t), \partial_t \hat \phi_{\bf p'}^\dagger 
(t)]= i \, \delta^3({\bf p}-{\bf p'}) \, .
\label{etc}
\ee
When decomposing the mode operator as
\be
\hat\phi_{\bf p}(t) = \hat a_{\bf p} \, \phi_p(t) + \hat a^\dagger_{-\bf p} \, \phi_p^*(t)\, ,
\ee
 where the destruction and creation
 operators satisfy the usual commutators
 \be
 [\hat a_{\bf p}, \hat a^\dagger_{\bf p'}]= \delta^3({\bf p}-{\bf p'})  
 \, , \quad [\hat a_{\bf p},  \hat a_{\bf p'}]=0 \, ,
 \label{uscom}
 \ee
eq. (\ref{etc}) 
is verified at all times 
because the Wronskian of the positive frequency
(c-number) mode \be
\phi_p(t) = e^{- i \om_p t}
/(2 \om_p)^{1/2}\, ,
\label{fm}
\ee
is constant (and conventionally taken to be unity). 

Had an odd term like $\gamma \partial_t$ been present in eq. (\ref{eqom})
the constancy of the Wronskian would have been lost.
Hence
the possibility of realizing the ETC (\ref{etc})
with the help of eq. (\ref{uscom}) would have been lost as well. 
This already indicates
that, unlike dispersive (real) effects,
dissipative effects
require more general settings than the above. 

\subsection{Interacting models breaking $LI$, general properties}

We now introduce additional 
degrees of freedom, 
here after collectively named $\Psi$,
which 
induce dissipation above the energy $\llv$. 
We shall work with a particular class of simple models 
in order to get an 
exact (non-perturbative) 
expression 
for the two-point function of $\phi$ of
eq. (\ref{Gw}). 
Before introducing these 
models,
we derive 
general results 
valid for {\it all} unitary QFT's 
possessing
dissipative effects above $\llv$ in the ground state (the interacting vacuum). 

We assume that the total action decomposes as
\be
S_T = S_\phi +  S_{{\Psi}} +  S_{\phi, {\Psi}} \, ,
\ee
where the first action is that of eq. (\ref{freeS}), the second one
governs the evolution of the ${\Psi}$ fields, and the last one
the coupling between $\phi$ and these new fields. 

We also impose that the last two actions (and the state of the system,
the density matrix
$\hat \rho_T$) 
preserve the homogeneity and isotropy of Minkowski space. 
From now on, the cartesian coordinates $t, {\bf x}$ 
are at rest with respect to the preferred frame
which is defined by the action $S_{{\Psi}} + S_{\phi, {\Psi}}$,
i.e. in a covariant notation $\partial_t \equiv l^\mu \partial_\mu$,
see eq. (\ref{perp}). 
For these models, 
the Fourier transform of the Wightman function of eq. (\ref{Gw})
is of the form
\ba
G_{\bf p, \bf p'}(t,t') 
 &=& {\rm Tr}\, [ \hat \rho_T \, \hat \phi_{\bf p}(t) \, \hat 
 \phi^\dagger_{\bf p'}(t')] \, ,
 \nonumber \\
&=& G_W(t,t'; p) \, \delta^3({\bf p}-{\bf p'}) \, .
\label{Gf}
\ea
At this point, an 
important remark should be made. 
In the presence of interactions, 
 the notion (and the usefulness) of the time-dependent 
 modes given in eq. (\ref{fm}) 
 disappears. 
Instead, 
the time-dependent 
 function $G_W(t,t'; p)$ of eq. (\ref{Gf}) is always 
 well-defined,  for "all" choices of ${\Psi}$ 
 and for all actions $S_{{\Psi}} +  S_{\phi, {\Psi}}$.  
 We shall return to this point in Appendix B after eq. (\ref{SfD}). 
 

When the situation
is stationary,
$ G_W(p; t,t')$ further simplifies in the frequency
representation:
\be
G_W(t,t'; p) = \int {d\om \over 2\pi} e^{-i\om(t-t')} G_W(\om, p) \, .
\label{FTdef}\ee
When working in the ground state, one reaches 
the simplest case 
in the following sense.
Whatever the ${\Psi}$
fields may be, the Fourier transform of the
time ordered (Feynman) propagator, 
is always of the form given in eq. (\ref{GfF}),
and therefore characterized by a single function, the
Feynman self-energy $\Sigma_F(\om,p)$. 
When restricting attention to Gaussian models,
$\Sigma_F(\om,p)$ is given by a 1-loop 
calculation, whereas it contains a series of 1PI graphs
for non-Gaussian models. 
In non-vacuum states and in non-stationary situations, 
2pt functions and self-energies have 
a more complicated structure~\cite{Arteaga:2007us}.
To 
understand 
this strucure it is useful to study separately  
the commutator 
 $G_c$ of $\phi_{\bf p}$, 
  which is odd in $\om$,
 and the anti-commutator $G_a$, which is even.
%

%

Let us conclude with two remarks.
First, the effective dispersion relation of $\phi$ 
is {\it a posteriori} defined by the poles of eq. (\ref{GfF}).
In this way, non-trivial dispersion relations arise from
dynamical processes rather than from being introduced from
the outset. The present work therefore provides physical 
foundations (and restrictions, as later discussed) 
to the kinematical approach which has been adopted in the literature. 
Second, from analyzing dynamical models, we shall see
that, even in the vacuum, 
{\it on-shell} dissipative effects (i.e. dissipation arising along the
minima of the denominator of eq. (\ref{GfF}))
are 
unavoidable
when $LI$ is broken in the UV by the action $S_{\Psi}+S_{\phi, \Psi}$, 
in complete opposition 
with the fact that 
on-shell dissipation is forbidden when 
working (in the vacuum) with  
$LI$ actions. 

\subsection{Gaussian models}

To simplify the calculation of $\Sigma(\om,p)$
and to get non perturbative expressions, 
we assume that the action $S_T$ 
is quadratic in all field variables.
At first sight this could be considered as an 
unjustified 
hypothesis. However,
it should be recalled that we are {not} 
after computing $\Sigma$ from first principles. Rather we aim to 
compute
the power spectra of (\ref{PandS}) 
{\it given} the properties of $\Sigma$, and this can be done
with quadratic models.


%
Since we are preserving the homogeneity of Minkowski space, 
the Gaussian assumption 
implies that 
the total action splits as 
\be
S_T = \int \! d^3p \, S_T({\bf p}) \, , 
\label{psep}
\ee
where each action $ S_T({\bf p})$
depends only on the {complex}
mode 
operators $\phi_{\bf p}$, $\Psi_i({\bf p})$. 
(Since
$\phi_{\bf p}^\dagger = \phi_{\bf - p}$, $\Psi_i^\dagger({\bf p})
= \Psi_i({\bf -p})$,
 $S_T({\bf p}) + S_T({\bf -p})$
is real.)
The structure of $S_T({\bf p})$ is 
\ba
 S_T({\bf p}) &=& {1 \over 2 } \int dt\ \phi_{\bf p}^* 
 (- \partial_t^2 -  \om_p^2 ) \phi_{\bf p} 
+ {1 \over 2 } \Sigma_i \int dt\ \Psi_i^*({\bf p})
 ( - \partial_t^2 - \Omega^2_i(p)) \Psi_i({\bf p})
\nonumber\\
&& + \Sigma_i \int dt\ g_i(p) \,  \phi_{\bf p} \, 
\Psi_i^*({\bf p}) \, ,
\label{Sd}
 \ea
 where $i$ is a discrete (or continuous) index, 
 where $\Omega_i(p)$ is the frequency 
 of the oscillators $\Psi_i({\bf p})$, 
 and where $g_i(p)$ is the coupling constant at fixed 
 $p, i$.\footnote{When 
 this action was presented in seminars, it was often found {\it ad hoc}
 and/or {\it artificial}. We dispute both charges for the following reasons. 
 Concerning the {\it ad hocery}, 
 it should be realized that {\it any}
 unitary (homogeneous)
 model exhibiting dissipation 
 will be
 governed by an action of the type (\ref{Sd}), either
 strictly speaking, or in the sense that the 2pt function of $\phi$, 
 eq. (\ref{Gw}), 
  will {\it exactly} behave as 
 that obtained from a model governed by an action of the type (\ref{Sd}).  
 Given this, the {\it artificial}
 qualification appears to us 
  also unjustified since the 2-pt functions derived from
  "physically well-founded" models 
 would perforce belong to the class derived from eq. (\ref{Sd}).
 It should be noticed that the {\it ad hoc} qualification has been 
 also often used against dispersive models, 
 apparantly not realizing that the robusteness of the spectra
 was established essentially for all 
 dispersion relations~\cite{Niemeyer:2000eh,Balbinot:2006ua}. 
 What was probably 
 meant by 
 is that it is not yet known which of these 
 relations is actually revelant, but this is another issue.}
%

Models of this type have been used 
since many years (at least since the early 50's) and
for different purposes. They have been introduced 
(in their continuous version) 
to study non-pertubatively atomic transitions
 and to model quantum electrodynamics in the dipole approximation, 
 see \cite{1976AnPhy..98..264A}
 and refs. therein, 
 see also \cite{Raine:1991kc,Massar:1900vg} for an application to
 the Unruh effect. They have been used in Quantum Optics~\cite{QNoise},
  to model quantum Brownian motion,
%
 and to study decoherence 
 effects~\cite{Unruh:1989dd}.
%
%
 Needless to say that
 our intention is not to provide a new approach to solve
 these models.%
 \footnote{
 Depending on the point of view adopted, they 
 can be
 solved and analyzed by means of different methods.
 In what follows we shall 
 use the simplest approach based on Heisenberg equations of motion. 
 Since we shall work at zero temperature, this approach is appropriate
 because quantum aspects dominate over stochastic ones. 
 Even though legitimate, we shall not use the general methods\cite{QNoise} 
 which have been developed to study 
"open quantum systems" since 
these 
 methods (Influence Functional, Master Equation, Closed Time Path Integral) 
  somehow 
 hide the simplicity of the models we are dealing with.
 In fact only but standard Quantum Mechanics is required to
 solve them.  Moreover 
 we are planning (in a subsequent work)
 to 
 study the correlations between $\phi$ and $\Psi_i$. Therefore we shall
 treat $\phi$ and $\Psi_i$ on equal footing 
 as in \cite{1976AnPhy..98..264A,Raine:1991kc,Massar:1900vg}.}

Instead of chosing a priori a specific model, 
we shall solve 
the equations of motion
 without specifying 
the set of $\Psi_i({\bf p})$,
their energy $\Omega_i(p)$ and their coupling $g_i(p)$. 
We shall choose them in the next subsection 
to further simplify
the equations we shall need to solve.
%
To preserve stationarity in Minkowski space, the $\Omega$'s  and  the $g$'s
must be time independent. 
 When 
 $\Omega_i^2(p) \neq M_i^2 + p^2$, the kinetic action of $\Psi_i({\bf p})$
 breaks 
 $LI$ and defines 
 the preferred frame. On the contrary when  $\Omega_i^2(p) = M_i^2 + p^2$
 the preferred frame is only defined by the interaction term $S_{\phi \Psi}$
  through the $p$-dependence of the coupling functions $g_i(p)$. 
%
 

  %
 The equations of motion are
 \ba
 &&( \partial_t^2 +  \omega_p^2 ) \, \phi_{\bf p} = \Sigma_i \,  g_i(p) \, 
 \Psi_i({\bf p}) \, ,
 \label{eqphi}
 \\
 && (  \partial_t^2 +  \Omega^2_i) \Psi_i({\bf p}) = 
 g_i(p) \, \phi_{\bf p} \, .
 \label{eqPsi}
 \ea
 The general solution of the second equation reads
 \be
  \Psi_i({\bf p}, t)=\Psi_i^o({\bf p}, t) + \int dt' R_i^o(t,t'; p) \,
  g_i(t'; p)  \, \phi_{\bf p}(t')\, ,
  \label{solPsi}
  \ee
  where $\Psi_i^o({\bf p}, t)$ is a free solution which depends
  on initial conditions imposed on $\Psi_i({\bf p})$. 
  The second term contains
  $R_i^o(t,t';p) $, the (free) retarded Green function 
  of $\Psi_i({\bf p})$. It obeys
  \be
  (  \partial_t^2 + \Omega^2_i(p)) \,  R_i^o(t,t'; p) = \delta(t-t')\, ,
  \ee
  and vanishes for $t < t'$. 
  (To prepare the application to time dependent geometries, 
  we have treated $\omega_p^2$, $\Omega_i^2$ and $g_i$ as arbitrary functions
  of time (in cosmology these 
   become indeed time
  dependent through their dependence in the scale factor $a(t)$). )
  Injecting eq. (\ref{solPsi}) in eq.  (\ref{eqphi}) one gets
  \be
  ( \partial_t^2 + \om_p^2 ) \phi_{\bf p} = 
  \Sigma_i \, g_i(t; p) 
  \Psi_i^o({\bf p}, t)
  + \Sigma_i   \, g_i(t; p) \int dt' 
  R_i^o(t,t'; p) 
   g_i(t'; p)  \phi_{\bf p}(t') \, .
  \label{eqtosimpl}\ee
  The (exact) 
  solution of this equation has always the following structure 
  \be
  \phi_{\bf p}(t) = \phi_{\bf p}^d(t) + \int dt' G_r(t,t'; p) 
  [\Sigma_i \,  g_i(t'; p)
  \Psi^o_i({\bf p}, t')] \, .
  \label{trues}
  \ee
  The first term is the ''decaying'' solution. It contains
  all the information about the initial condition of $\phi_{\bf p}$, and obeys
  the non local equation
  \be
  \int\!dt_1 [ \, \delta(t-t_1)(\partial_{t_1}^2 + \om_p^2) 
  - \Sigma_i  \, g_i(t; p) \,
  R_i^o(t,t_1; p) \,
  g_i(t_1; p) ]\phi_{\bf p}(t_1) = 0\, . 
  \label{Gr}
  \ee
  The second term is the ''driven'' solution.
  It is governed by the initial conditions of $\Psi_i^o({\bf p})$
  and by the (dressed) retarded Green function, the solution of
  eq. (\ref{Gr}) with $\delta(t-t_1)$ on the r.h.s.
%
  Therefore the evolution of both $\phi^d$ and $G_r$
  fully takes into account, through the non-local term in the 
  above bracket, the backreaction due to the coupling
  to the additional degrees of freedom. In Gaussian models,
  the backreaction is quadratic in $g_i$. 
  Hence the solutions $\phi^d$ and $G_r$ 
  are series containing all even powers of $g_i$. 
  Therefore Gaussian models do give rise
  to non perturbative effects. Moreover 
  since $g_i(t;p)$ are arbitrary functions of $p$ and $t$,  at this point 
  there is no reason to consider non-Gaussian models.
 
  We conclude this subsection with 
 two 
 remarks.
   First, eq. (\ref{trues}) {also} furnishes the exact solution
  for the Heisenberg operator $\phi_{\bf p}$
  because the equations we solved are all linear.
  Since we work quantum mechanically,
  it is 
  necessary  to study 
  the 2pt 
  functions of $ \phi_{\bf p}$.
  We review their properties in the Appendices.
  
   Second,
  eq. (\ref{Gr}) tells us that in general, the coupling
   to an environment (even a linear one) gives rise to {\it non-local}
   equations of motion. When dealing with stationary situations
   this does not cause any problem because, as shown in Appendix B,
    observables can be (algebraically)
   computed in the frequency representation. 
   However in non stationary situations and in curved space-times, 
   to be able to compute observables
   it becomes imperative to simplify  eq. (\ref{Gr}). 
   A question we now address.
%

 \subsection{
Time dependent settings}

In Appendix C, we provide a class of 
 stationary models characterized by the power of the ratio $p/\llv$
which specifies how dissipative effects grow with the energy, see
eqs. (\ref{ndiss}, \ref{iDRn2}).
This class covers the general case and 
can be used as a template to study the phenomonological
consequences of dissipative effects. 
In addition, after eq. (\ref{DSr}), it was noticed that the dissipative properties 
are governed by the product 
$ g^2(\om,p) R^o(\om,p)$, as can be understood from 
eq. (\ref{Gr}). Therefore
theories with different degrees of freedom $\Psi$ and with 
different coupling terms but delivering  the same product
will give rise to the same (stationary) phenomenology.

In this Section we exploit 
this freedom to simplify the expressions 
having in mind 
the transposition of our model from Minkowski space to 
 curved 
  metrics. 
 Therefore 
the models we shall use 
should possess two-point functions with
 simple properties when expressed in the terms of space-time coordinates 
 (and not in Fourier
 components $\om,p$).  
  
  The core of the problem is that, when dissipation is strong, 
  the relevant observable $G_a$, see eq. (\ref{PandS}),
 is given by the "driven" term in eq. (\ref{Gags}), which is 
  a double integral containing a kernel (which can be 
  computed) and the retarded Green function of $\phi$ which is not
 known. Indeed it is only implicitely defined as a solution of 
 eq. (\ref{Gr}), which is, in general, 
 {non-local}. 
 We are thus led to choose the action of $\Psi$ 
 in order for  eq. (\ref{Gr}) 
 to be local. 
 Notice that this requirement
  concerns the Green function of the $\Psi$-environment
  and not that of the $\phi$-system we probe. Thus it will not restrict the 
  latter's phenomenology.

 Given this choice, an appropriate 
 class of models 
 is defined by the action 
 \ba
   S_T^{(n)}({\bf p}) &=& {1 \over 2 } \int\!dt\ \phi_{\bf p}^* 
 (- \partial_t^2 -  \om_p^2 ) \phi_{\bf p} 
\nonumber\\
&& + {1 \over 2 }  \int\!dt\int\!dk\ \Psi^* ({\bf p},k) 
 ( - \partial_t^2 - \left({\pi\llv k}\right)^2 
 ) \Psi({\bf p}, k)
\nonumber\\
&& + {g \llv} \int\! dt\int\! dk  \, 
\left({p \over \llv}\right)^{n+1} 
\phi_{\bf p} \, 
\partial_t \Psi^* ({\bf p},k) \, .
 \label{dynM}
 \ea
When compared with the action of
eq. (\ref{Sd}), we have replaced the discrete index $i$ 
by the integral over the dimensionless variable $k$. 
As recalled in Appendix B, one needs a continuous spectrum of the
environment to have proper dissipation, see discussion after eq. (\ref{DSr}).
The continuous variable $k$ can be viewed 
as a momentum 
(expressed in units of $\llv$) in an extra flat fifth spatial dimension. 
The relationship with the Brane 
World Scenarios of~\cite{Libanov:2005yf,Libanov:2005nv} is clear. 
In the 'atomic' version of 
this model~\cite{Raine:1991kc,Massar:1900vg}
 which has inspired us, 
the radiation field $\Psi$ is a massless 2 dimensional field 
propagating in $t$ and in the dimension associated with $k$.

In $S_{\phi \Psi}$ we have factorized out $\llv$ so that the 
coupling constant $g$ is dimensionless.  
We have also introduced an additional time derivative 
acting on $\Psi$. 
This choice leads to the above mentioned $\delta(t-t')$.
Indeed, on the one hand, taking this extra derivative, the continuous
character of $k$, and the fact that $g$ is independent of $k$,
 eq. (\ref{eqtosimpl}) becomes
 \be
 ( \partial_t^2 + \om_p^2 ) \phi_{\bf p} = 
   g_n \, 
  \partial_t  \int\! dk   \Psi^o({\bf p},k, t)
   -  \, g_n  \,  \partial_t  \int dt' 
   \int\! dk   R^o(t,t'; k, p) \, 
   \partial_{t'} \big( g_n \phi_{\bf p}(t')\big) \, ,
 \label{eqsimpl}
 \ee
 where $g_n \equiv g \llv \left({p / \llv}\right)^{n+1} $.
On the other hand, for each 3-momentum ${\bf p}$, 
the Green function of  ${\Xi}= \int dk \Psi(k)$ 
 is that of a massless 2-dimensional free field. 
In Fourier it is given by $R^o(\om, k)= 1/(-(\om+ i \e)^2 + ({\pi\llv k } )^2)$. Hence it obeys
 \be
\partial_t {\bf R}^o(t,t') \equiv
\partial_t
\int\! {d\om \over 2 \pi} \int\! dk \, R^o(\om, k)\,  e^{-i \om(t-t')} 
= 
{\delta(t-t')
\over \llv}, 
\label{deltat} 
\ee
which is the required property to simplify eq. (\ref{eqsimpl}).

When $g_n$ is constant,
the retarded Green function of $\phi$ associated to eq. (\ref{eqsimpl})
obeys the following {local} equation 
\be
 [    \partial_t^2 +   {g_n^2\over \llv} 
  \partial_t + \om_p^2     ] \, G_r(t,t', p) = \delta(t-t')\, ,
 \ee
 To make contact with Appendix B and C, let us rewrite this equation
 in Fourier transform, 
 \be
 [  - \om^2 - ig^2   {\om \over \llv}\,  p^2 \big( { p \over \llv}\big)^{2n}
  + \om_p^2     ] \, G_r(\om, p) = 1 \, .
 \ee
 We thus see that Re$\Sigma_r=0$ and that Im$\Sigma_r$ is 
 (exactly) given by $g^2 $ times the expression of  eq. (\ref{iDRn2}).
Thus, even though we have chosen a simple 
and definite form for ${\bf R}^o(t,t')$, 
the above action does
deliver the $n$ dissipative behaviors of Appendix B
by choosing the appropriate power of $p$ in the coupling $g_n$ 
appearing in the action $S_{\phi \Psi}$. 
%
%

  When $g$ and $\om_p^2$ are arbitrary
 time-dependent functions, the Fourier analysis looses its power.
 However, in time dependent settings, 
 our 
 $G_r$ still obeys a local equation: 
 \be
 \left[    \partial_t^2 + 2\tilde \gamma_n  
  \partial_t + [ \om_p^2 + \partial_t  \tilde \gamma_n
  ] \right] 
  G_r(t,t') = \delta(t-t') \, ,
 \label{tdefGr}
 \ee
 where the $n$-th decay rate $\tilde \gamma_n(t) = g^2(t) \gamma_n$ 
 is now a definite time dependent function. 
 When transposing 
 the model of eq. (\ref{dynM})
  in expanding homogeneous
  universes, the Green function will therefore also obey a local equation.

 \subsection{Covariant description} 
 
 We now provide the covariant version of the action of eq. (\ref{dynM}).
This expression will then be 
used 
to define our theory 
 in curved backgrounds.\footnote{Perhaps the motivations for these two
 steps require further explanation since another referee wrote: {\it 
 "I do not see the point of a covariant
 description of the $\Psi$ field since it is nothing more than a convenient
 parametrization of the environment degrees of freedom."} 
 Several points should
 be mentionned. First, 
as in studies of dispersive effects, we require that our QFT 
  be well defined in arbitrary backgrounds. It would be physically
  questionable (or even meaningless) 
  to study dissipative effects on the observables of eq. (\ref{PandS})
  from theories which can be
  defined only in a particular class of space-times.
Second, 
we have introduced the $\Psi$ field to parametrize 
  the effects of dissipation which would stem not from any theory, but only
  from fundamental theories
  obeying the Equivalence Principle, 
  see \cite{Libanov:2005yf} for a prototype.  
  Third, 
  from the point of view of open quantum systems, it could indeed
seem inappropriate to proceed to a covariantization, since 
 in most 
  situations (heat bath, condensed matter
 systems) there is a preferred frame which is globally defined. 
 However, 
 when the system is non-homogeneous (e.g. a fluid
  characterized by a non-homogeneous flow), 
 low energy fluctuations effectively live in a 
 curved geometry\cite{Unruh:1980cg}.
 Moreover, in this case, 
 short distance effects, i.e. dispersive (or dissipative) 
 effects, are 
 covariantly described 
 when using this metric~\cite{Unruh:1994je,Jacobson:1996zs}, 
 because they arise {\it locally}. 
 In brief,
 we shall 
 write down actions in curved space using
 the EP which requires a covariant description.}
 
 Two steps should be done. We need to go from $\bf p$ considerations 
 to a local description, and express the various actions in terms of 
 the unit vector field $l^\mu$ and the spatial metric $\perp^{\mu \nu}$
 defined in eq. (\ref{perp}). 
 Both are straightforward and, in arbitrary coordinates, the total action
  $S^{(n)}_T = \int d^3p \, S_T^{(n)}({\bf p})$
  reads (in Minkowski space-time)
 \ba
   S_T^{(n)}&=& - {1 \over 2 } \int\!d^4x \sqrt{-g}\,  g^{\mu \nu}
   \partial_\mu \phi \partial_\nu \phi
\nonumber\\
&&\,  + {1 \over 2 } \int\!d^4x \sqrt{-g} 
 \int\!dk\
 \Big[ \big(  l^\mu l^\nu - 
 c^2_\Psi \!\perp^{\mu \nu} \big) 
\partial_\mu  \Psi_k  
 \partial_\nu\Psi_k
  - \left({\pi\llv k}\right)^2 
  \Psi^2_k \big) \Big]
\nonumber\\
&& +   { g \llv}\int\!d^4x \sqrt{-g} 
\Big( \big({ \Delta 
\over \llv^2 }\big)^{(n+1)/2} 
\phi \Big)  \, 
l^\mu \partial_\mu \int\! dk \Psi _k \, ,
 \label{CdynM}
 \ea
where 
$\Delta$ is the Laplacian on the three surfaces
orthogonal to $l^\mu$.
 
 We have slightly generalized the action of eq. (\ref{dynM})
 by subtracting 
  $c^2_\Psi\!\perp^{\mu \nu} $ to the kinetic term of $\Psi$,
 where $c^2_\Psi \ll 1$.
 With this new term, $p$-th components of the field $\Psi_k$ 
 are now massive fields which propagate 
 with a velocity (relative to $l^\mu$) whose square is bounded by $c^2_\Psi$.
 In addition they now possess a well defined energy-momentum
 tensor which can be obtained by varying their action with respect to
 $g^{\mu \nu}$. To obtain 
 the simplified expressions we use in the paper 
 the (regular) limit $c^2_\Psi \to 0$ should be taken.

 In this limit, the (free) retarded Green function 
 of the ${\Xi} = \int dk \Psi$ field 
 obeys a 
 particularly simple equation 
 when expressed in space-time coordinates. Indeed, it obeys
 \be
 l^\mu {\partial \over \partial x^\mu} \int dk R^o(x,y;k)  \equiv 
 l^\mu {\partial \over \partial x^\mu} {\bf  R}^o(x,y) ={ 
 1\over \llv}
  {\delta^4(x^\mu -y^\mu)\over \sqrt{-g}}\, .
  \label{covdelta}
 \ee
 On the r.h.s, one finds 
 the "invariant" delta function
 with respect to the invariant measure $d^4x \sqrt{-g}$. 
 This equation is nothing but the covariantized and "localized" version
 of eq. (\ref{deltat}).
 Its physical meaning is clear: 
 the back-reaction on $\phi(x)$ 
 through ${\Xi}$ is strictly local.
 
 Nevertheless, the equations of motions 
 do not have a particularly simple form
in space-time coordinates. 
 Using the condensed notation ${\Xi} = \int dk \Psi$,
  the relevant equations read 
 \be
 { 1 \over \sqrt{-g}}\, \partial_\mu \sqrt{-g} \, g^{\mu \nu} \partial_\nu \, 
\phi(x) =  g\llv {  1\over \sqrt{-g}} 
 \big({ \Delta \over \llv^2 }\big)^{n+1 \over 2} \sqrt{-g}\, 
  l^\mu\partial_\mu {\Xi}(x) \, ,
\label{coveomphi}
\ee
where the interacting field ${\Xi}$ field is 
\be
{\Xi}(x)= {\Xi}^o(x) 
-  \, g\llv 
\int\! d^4y \sqrt{-g}\,  {\bf R}^o(x,y) \Big[ { 1 \over \sqrt{-g}}\, 
{\partial_\mu }
\Big(l^\mu   \sqrt{-g}  \big({ \Delta \over \llv^2 }\big)^{n+1 \over 2} \phi(y)
\Big)\Big] \, .
\label{covpsisol}
\ee
 When inserting eq. (\ref{covpsisol}) in eq. (\ref{coveomphi}), 
 using eq. (\ref{covdelta}), one verifies that the dissipative term
is local and first order in $l^\mu \partial_\mu$.
Dissipation thus occurs along the preferred direction.
 
\section{Dissipative effects in curved space-time} 
\label{curvedbg}

 \subsection{The Lagrangian and the properties of $\Psi$}

%
%
 
 To define a dissipative QFT in an arbitrary curved geometry, 
 we need some principles. 
 From a physical point of view, we adopt the Equivalence Principle,
 or better what can be considered as its generalization in the presence
 of the 
 unit time-like vector field. We are in fact dealing
 with two (sets of) dynamical fields, the $\phi$ field we probe,
 and the ${\Psi}_i$ fields we do not;  
 but also with two background fields $g_{\mu \nu}$ and $l^\mu$.  
 The Generalized Equivalence Principle means that the action densities of the 
 dynamical fields be given by scalar functions (under general coordinate 
 transformations) which reduce 
 to those one had in Minkowski space time and for a homogeneous and static
 $l^\mu$ field, i.e. those of eq. (\ref{CdynM}).
 
 The densities are thus not completely fixed by the GEP, 
 as it was the case with the EP.
 For a scalar field in a curved geometry, there was always the 
 possibility of considering a non-minimal coupling to gravity
 by 
 adding to the lagrangian 
 a term proportional to $R \phi^2$. 
  In the present case, the ambiguity is larger because
  the vector $l^\mu$ allows to form
 new scalars, 
 such as its {expansion} 
 \be
 \Theta \equiv \nabla_\mu l^\mu = (-g)^{-1/2}\partial_\mu[( -g  )^{1/2}l^\mu]
 \, ,
 \label{exp}
 \ee 
 where 
 $\nabla_\mu $ is
 the covariant derivative with respect to the metric $g_{\mu\nu}$.
 The ambiguity can only be resolved by 
adopting some additional principle, such as 
the principle of minimal couplings which forbids 
adding such scalars. 

Rather than adopting it, our second principle (which simplifies the equations
of motion, 
but which is by no means necessary) is that the 
{\it locality} of the back-reaction effects of $\phi$ through $\Psi$
be preserved. That is, we choose the non-minimal coupling of $\Psi$
so as to keep eq. (\ref{covdelta}). 
Starting from eq. (\ref{CdynM}), 
the locality is preserved by replacing 
both in $S_\Psi$ and $S_{\Psi \phi}$ 
 \ba
l^\mu \partial_\mu\Psi_k \to 
{\cal D}_l \Psi_k \equiv 
l^\mu \partial_\mu\Psi_k +{ \Theta  \over 2}\,  \Psi_k
= {1 \over 2}   \left(  l^\mu
\nabla_\mu \Psi_k +  \nabla_\mu \, [l^\mu \Psi_k] \right)\, .
\label{covl}
 \ea

To simplify the forthcoming equations, we use the fact
 that one can always work in "preferred" coordinate systems 
 in which the shift $l^i$ vanishes and in which 
 the preferred time is such that $l_0 = 1$. 
 (We assume that the set of orbits of $l^\mu$ is complete 
 and without caustic. In this case, every point of the manifold
 is reached by one orbit.)
It is useful to work in these coordinate systems
with rescaled fields $\Psi^r \equiv (-g)^{1/4}\,
 \Psi$ because one can then group the above two terms into a single expression:
\be
{\cal D}_l \Psi_k
= (-g)^{-1/4} \, l^\mu  
\partial_\mu ( (-g)^{ 1/4}\,\Psi_k)= (-g)^{-1/4} \, l^\mu  
\partial_\mu \Psi_{k}^r \, .
\label{covl2}
 \ee
%
%
 Notice also that there exists a 
 subclass of 
 background fields $(g,l)$, for which one can find 
  coordinate systems such that
 {\it both} the shift $l^i$ and $g^{oi}$ vanish.
 In these {\it comoving} coordinate systems, 
 the above equations further simplify  since only the spatial part of the 
 metric matters because
 $ -g=  h_c$ where 
 $h \equiv {\rm det}(\perp_{ij})$.
  ~\footnote{\label{intfnote} 
 To illustrate this point, 
 let us consider FLRW flat metrics. In comoving coordinates, one 
 has $ds^2 = -dt^2 + a^2 dx^2$, and $h=h_c= a^6$. In Lema\^itre coordinates
 $X=a \, x$,  one has $ds^2 = -dt^2 +(dX - V dt)^2$, 
 where the velocity is 
 $V=HX$. 
 The spatial sections are now the Euclidean space with $h=1$, and 
 the (contravariant) components of the unit vector field are $l^\mu= 1, V$. 
 To compute $h_c$ one should solve the equation of motion of comoving 
 (free falling) 
 observers $dX -V dt = 0$, and use the initial position
 as new coordinates. 
 This procedure will be 
 done in subsection \ref{DBHm} starting with Painlev\'e-Gullstrand 
 coordinates to describe the
 black hole metric and $l^\mu$ field.} 
%
%

Having chosen this non-minimal coupling, 
one verifies that the kinetic term 
of the rescaled fields $\Psi_r$ 
is insensitive to the "curvature" of both $g_{\mu \nu}$
and $l^\mu$ (when the limit $c^2_\Psi \to 0$ is taken).
Moreover the differential operator which acts on the
retarded Green function of $\Psi_r$ in the equation 
of motion of $\phi$, see eqs. (\ref{coveomphi}, \ref{covpsisol}), is also "flat"
thereby guaranteeing that the modified version
eq. (\ref{covdelta}) still applies, that is
 \ba
 {\cal D}_l \, 
{\bf  R}^o(x,y)&=&    (-g(x))^{-1/4}   \, 
 l^\mu \partial_\mu \Big({\bf  R}_r^o(x,y)\Big) \,  (-g(y))^{-1/4}
  \nonumber \\
&=& {1 \over \llv} { \delta^4(x^\mu -y^\mu)
  \over \sqrt{-g}}\, ,
 \label{covcurveddelta}
 \ea
where ${\bf  R}_r^o(x,y)$
is the retarded Green function of the rescaled field $\Psi_r$.
It obeys (in preferred coordinate systems) 
$\partial_t {\bf  R}_r^o = \delta^4/ \llv$,
and "defines" the retarded Green function 
${\bf  R}^o= (-g)^{1/4}\,{\bf  R}_r^o \, (-g)^{1/4}$ which is a bi-scalar.
Hence the equations of motion in an arbitrary background "tensor-vector metric" specified by the couple ($g_{\mu\nu}, l^{\mu}$), are given by eqs. 
(\ref{coveomphi}, \ref{covpsisol}) 
with the substitution of eq. (\ref{covl}).

Several remarks should be made.
First, from the simplified 
equation $\partial_t {\bf  R}_r^o =\delta^4/ \llv$
it might seem that the background tensor metric $g_{\mu \nu}$ plays no role.
This is not true, it enters indirectly
as it is used to normalize the vector field $l^\mu$ at every point. 
Therefore, 
it intervenes 
in the specification that the (proper)
energy scale $\llv$ stays fixed as the universe (or the comoving volume
$h_c$) expands (as required by the Equivalence Principle).

Second, in the limit $c^2_\Psi \to 0$,
the (rescaled) $\Psi_k$ fields define a new kind of field. 
They propagate
in an 
effective space-time given by the time development
of the 3-dimensional set of orbits of the $l^\mu$ field. 
Indeed, at fixed $k$, $\Psi_k(x)$ can be decomposed in 
non-interacting local oscillators, each of them evolving separately
along its orbit. This situation is similar to the long wave length (gradient-free) limit of cosmological perturbations considered 
by Steward and Salopek\cite{Salopek:1992kk}.
  In the absence 
of $l^\mu$, the space time geometry must be (nearly) homogeneous for the 
 mode action to possess this decomposition. However,  when 
 the $l^\mu$ field is given, 
 one can identify, even in non-homogeneous metrics, 
every point in space-time in an invariant way by the spatial position of the
corresponding 
$l^\mu$-orbit at some time, and the proper time along the orbit (as long
as $l^\mu$ has no caustic). 
We can thus build covariant actions exploiting this possibility
and consider fields composed of a dense set of
 local oscillators (i.e. one has 1 degree of freedom per point)
at rest with respect to $l^\mu$.
The fields $\Psi_k$ belong to this class of fields, here
after called {\it Chelva} fields.

Third, the analogy between the above formalism
and the DeWitt~\cite{DeWitt:1975ys}
  way to handle the short distance
behavior of Green (Hadamard) functions seems worth deepening.

Fourth, even though we have chosen to simplify the equation of motion of $\phi$
by imposing that the rescaled kernel obeys
$\partial_t {\bf  R}_r^o =\delta^4/ \llv$,
the 2pt functions of $\phi$ in curved backgrounds are highly non trivial,
and in particular the anti-commutator $G_{a}$.
Indeed $G_{a}$ is given 
by a double time integral, see eq. (\ref{Gags}), governed 
by the kernel ${\bf N}$ which 
 cannot be approximated by a delta function in the vacuum, see Section 3
 in \cite{Parentani:2007dw}.  

%

%
%
%

 
 \subsection{Application to cosmology}
 To get the equation of motion 
 we consider eq. (\ref{CdynM}) (with the curved metric modifications 
 introduced in the former subsection) in a FLRW metric 
$ds^2 = - dt^2 + a^2(t) \, d{\bf x} d{\bf x}$. 
To simplify the notations, we use
 the conformal time $d\eta = dt/a $ and work with the 
 rescaled fields $\phi_r= a \phi$ and
$\Psi_r(k) = a^{3/2}\Psi(k)$. 
 Notice that their power in $a$ differs. 
 Dropping these $r$ indices, working in Fourier transform 
 with respect to the (dimensionless) comoving coordinates $\bf x$, 
  the equation of motion of Heisenberg operator $\phi_{\bf p}$ is 
  \be
  \big(\partial_\eta^2 + 
  2  \gamma_n\,  \partial_\eta  + ( \omega^2_p(\eta) 
  + \partial_\eta\gamma_n)
  \big)  \, \phi_{\bf p} =  
  g_n \, \partial_\eta {\Xi}^o({\bf p})\, ,
  \label{cosmoeom} 
  \ee
  The conformal frequency
  $\omega^2_p(\eta)= p^2 - 
  \partial^2_\eta a/ a$ is that 
  of a rescaled 
  minimal coupled massless field. 
  In this expression, as everywhere in this subsection,
  $p$ is  the conformal (dimensionless and constant) wave vector. 
  The time dependent coupling coefficient 
  is \be
  g_n \equiv g\  a^{1/2}
   \ \llv \left({p / a \llv}\right)^{n+1} \, .
   \ee 
Its unusual $a$-dependence follows from 
    the different rescaling of the two fields.
 Straightforward algebra gives $\gamma_n(\eta)$,
   the decay rate in conformal time.
When compared to the
  comoving frequency $p$ to get the relative strength of dissipation, one gets
  \be
   {\gamma_n(\eta)\over p} = {1 \over 2}
    \big({g_n^2 \over p \llv }\big)
   = {1 \over 2} \big({p \over a \llv}\big)^{2n+1} = 
   {1 \over 2} \big({p_{phys}(a) \over \llv}\big)^{2n+1}\, . 
   \label{gammaninfl}
  \ee 
  In the last equality we have re-introduced 
  the proper momentum $p_{phys}(a) = p/a $.
  With this equation we verify that, at any time in an expanding universe
  and for any mode $\phi_p$,
  the relative strength of dissipation is simply that 
   obtained in Minkowski space at the corresponding energy scale, see 
   eq. (\ref{decayn}).
  This is nothing but the expression of the 
  Generalized Equivalence Principle, 
  at the level of the dynamical equations.
  
  Notice however that the equation of motion in an expanding
  universe contains a frequency shift
  \be
  \partial_\eta \gamma_n 
  = - ( 2n + 1) \, p^2  \, \big({aH \over p}\big)\, 
   \big({p \over a \llv}\big)^{2n+1} \, .
   \label{levels} \ee 
  We have factorized out $p^2$ 
  to get the
  relative value of the shift. It vanishes both when dissipation is negligible
  and when $aH/ p \ll 1$, i.e. when the expansion rate $H$ is negligible
  with respect to the proper momentum. From this expression
  we can already conclude that it cannot play any role
  when the two relevant scales $H$ and $\llv$ are well separated, 
  i.e. when 
  \be
  \sigma \equiv { H \over \llv }  \ll 1 \, .
    \label{psigma}
    \ee
   Indeed when the physical momentum is high and of the order of $\llv$, 
   the relative frequency shift is proportional to $\sigma$, and
   when the physical momentum is of the order of $H$ (at horizon exit, 
   see the Figure), it is  proportional to $\sigma^{2n+1}$. 
   \begin{center}
\includegraphics[width=8cm]{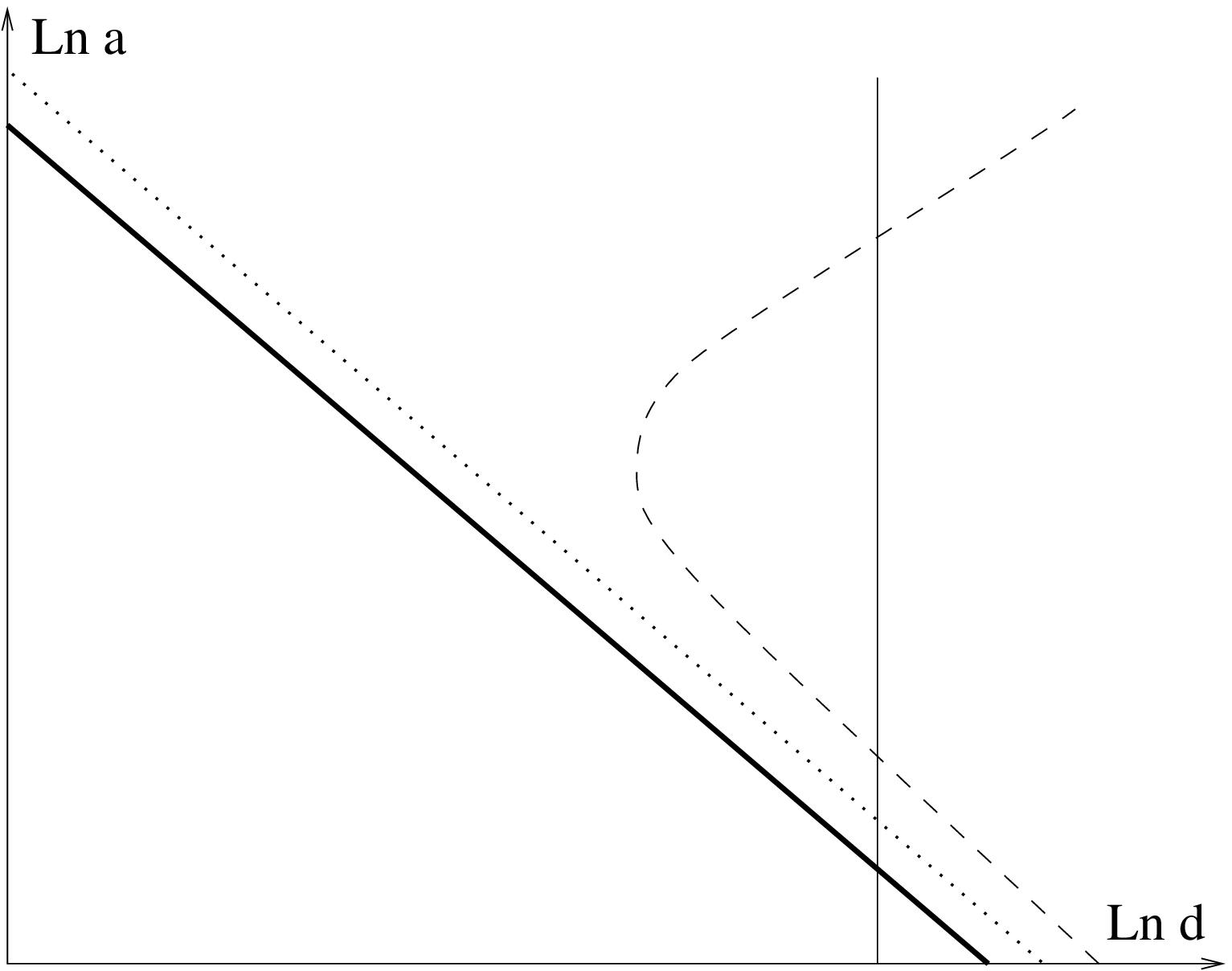}
\end{center}Figure caption.
{\it 
We have represented in a log-log plot and by a dashed line 
the evolution of $d_H = R_H/a$, the Hubble radius in comoving 
coordinates, as a function of $a$,
 both during inflation wherein $d_H \propto 1/ a$ and during the
radiation era wherein $d_H \propto a$. We have represented by a thick line
the trajectory of the cutoff length 
scale $d_\Lambda = 1/a\llv$ in the case where $1/\llv \ll R_H$ during inflation. 
The dotted line corresponds to an intermediate 
fixed proper length $\lambda$ which obeys
$1/\llv \!\ll\!
\lambda \!\ll\! R_H^{infl.}$.
The vertical line represents a fixed comoving scale $d_p = 1/p$. Below the 
cutoff length, all modes are overdamped. When a mode exits the cutoff length,
it becomes underdamped and starts propagating. When it crosses the intermediate 
length $\lambda$, it behaves as a free mode, and gets amplified only when
exiting the Hubble radius. As explained in the text,
adiabaticity, which is guaranteed by $1/\llv \ll R_H$, 
guarantees in turn that, near $\lambda$, modes are all born in the Bunch-Davies vacuum when the environment $\Psi$ is 
in its ground state.}
\\
%

Let us now establish that the power spectrum is robust, 
i.e. in spite of strong dissipative effects,
 when eq. (\ref{psigma}) is satisfied,
 the standard expression is recovered when the $\Psi$ 
 fields are in their ground state.

In free settings, when assuming that all $\phi_{\bf p}$ 
are in their ground state 
at the onset of inflation, i.e. working in the  Bunch-Davies vacuum,
the anti-commutator, see eq. (\ref{Gags0}), 
evaluated at equal time $\eta$ is simply
given by
\be
  G^{free}_a(\eta,p)
  =  \vert \phi^{in}_{\bf p} \vert^2 = {1 \over (2 p)} 
 \left( 1 + {1 \over (p\eta)^2 } \right)\, .
 \ee
In the last equality we worked for simplicity with a constant Hubble
parameter $H$, in which case $a = -1/H \eta$. 
Then the power spectrum of the 
physical (un-rescaled) field 
given by 
\be
P^{free}_p(\eta) =  2 p^3 {G^{free}_a(\eta,p) \over a^2(\eta)} 
= { H_p^2 
} \, (1 + (p\eta)^2) \, ,
 \ee
becomes constant after horizon exit.
We have added a $p$ subscript to $H$ because in slow roll inflation,
the relevant value of $H(t)$ for the $p$-mode is that evaluated
at horizon exit, i.e. $H_p = H(t_p)$, where $t_p$ is given by
$p/a = H$. The above equation shows that $P_p$ acquires some 
scale dependence through $H_p$. Similarly the
{\it deviations} from this standard behavior stemming from some
UV modification of the theory will also 
depend on $p$ through $H_p$ (and its derivatives).
For an explicit example, we
 refer to \cite{Niemeyer:2002kh} 
 where the modifications of the spectrum 
stem from the fact that the adiabatic vacuum is
imposed when $p/a_{in}$ is taken large but finite. 


In the presence of dissipation, the expression for 
$G_a$ 
radically differs from the above. 
Indeed, when dissipative effects grow with
the energy (as we suppose it is the case), 
one reaches the conclusion that in inflation
the decaying solution of eq. (\ref{cosmoeom})
is completely erased (unless one fine-tunes the number of e-foldings
so as to keep a residual amplitude). That is, the mode operator is entirely
given by its driven term, the second term in eq. (\ref{trues}). 
Hence the power spectrum is also 
purely driven 
and given 
by the second term of eq. (\ref{Gags}):
\be
G_a^{driven}(\eta,\eta,p) = \int\! d\eta_1 \int \!d\eta_2 \,
G_r(\eta, \eta_1, p)\,  G_r(\eta, \eta_2, p) \, {\bf N}(\eta_1, \eta_2, p)\, ,
\label{Gadriven}
\ee
where $G_r$ 
is the retarded Green function,
solution of eq. (\ref{cosmoeom})
with $\delta(\eta - \eta_1)$ as a source, 
and where the kernel ${\bf N}$ is the anti-commutator 
of $g_n \, l^\mu\partial_\mu {\Psi}_p$, the source of eq. (\ref{cosmoeom}).
From eq. (\ref{Gadriven}) we learn
that only the (initial) quantum state of the 
environment matters. In other words, 
because of the strong dissipation at early times,
the power spectrum is independent of
the initial state of $\phi$, whatever it was. 

In spite of these differences, 
when eq. (\ref{psigma}) is satisfied
and when the environment is in its ground state, the predictions
are unchanged, i.e. the power spectrum obtained from 
$G_a^{driven}$ coincides with that obtained with $G_a^{free}$
because the combined evolution of $\phi + \Psi$
consists in a 
(adiabatic) succession of 
stationary states ordered by the scale factor $a$.  
%
%

The proof goes in two steps.
First scale separation, 
guarantees the adiabaticity of the evolution.
That is, when $\eta$ and $\eta'$
are close (in the sense that $1 - a(\eta)/a(\eta') \ll 1$), 
$G_a^{driven}$
is well approximated by 
\be
G_a^{driven}(\eta,\eta', p) 
\simeq
 G_a^{statio}(\eta-\eta'; \omega_p(a),g_p(a)) = \int {d\om \over 2 \pi }\, 
e^{i \om (\eta-\eta')}\, 
 G_a(\omega ; \omega_p(a),g_p(a)) \, ,
\label{Gadiab}
\ee
where $ G_a(\om ;\omega_p(a),g_p(a))$ is the Fourier
 component 
of the anti-commutator calculated 
with the Hamiltonian characterized by the constant 
values of the frequency $\omega_p(a)$ and the coupling $g_p(a)$
both evaluated with $a= a(\eta)$.
In Appendix B, the stationary value of these anti-commutators
has been algebraically solved for all frequencies and all couplings, 
see eq. (\ref{Gaf}). 
Moreover, since by hypothesis, we are in the vacuum, 
eq. (\ref{fd}) also applies with $n(\om)=0$. 
In other words, the value of the anti-commutator follows from that of  
$G_c$.
This is enough as it guarantees that when the mode $\phi_p$ becomes
free, i.e. much after $\llv$-exit but before horizon-exit ($H \ll p/a \ll \llv$),
eq. (\ref{mie}) applies, see the Figure. 
Thus, {\it irrespectively} of what the coupling
with the environment was, adiabaticity implies
\be
G_a^{driven}(\eta,\eta, p) \to G_a^{free}(\eta,\eta, p) = {1 \over 2 \om_p(a)}\, ,
\ee
thereby guaranteeing that no modification of the power spectrum 
will be found.
In brief 
adiabaticity means that the evolution proceeds slow enough
for not inducing any (non-adiabatic) transition,
which in the present case would 
correspond to creation of pair of $\phi$ quanta with opposite wave vector $\bf p$. 
%
%
%
%
%
%
%

The second part of the proof consists in providing an upper bound
for the probability amplitude of obtaining a
non-adiabatic transition. This amplitude is
governed by the relative frequency change
\be
{\partial_\eta \om_p^{eff} \over (\om_p^{eff})^2 } 
= - {\gamma_n^2 \partial_\eta \ln \gamma_n \over (p^2 - \gamma_n^2)^{3/2}}
= (2n + 1) {H \, \gamma_{phys}^2  \over (p_{phys}^2 - \gamma_{phys}^2)^{3/2}} \, ,
\ee
where $p_{phys}= p/a$ and $\gamma_{phys}= \gamma/a$ 
are the physical momentum and decay rate.
Therefore, going backwards in time from the free regime
in the underdamped regime,
up to $p_{phys}^2 > 4 \gamma_{phys}^2$ (i.e. $p_{phys} <  \llv$), 
the non adiabatic parameter raises from zero 
but stays bounded by 
\be
{\partial_\eta \om_p^{eff} \over (\om_p^{eff})^2 } 
<  3 (2n + 1)\,  
\sigma \, \Big( {p_{phys} \over \llv}\Big)^{4n +1} < 3 (2n + 1)\,  \sigma \ll 1 \, .
\ee
This guarantees that the amplitude for 
the system to jump out  the ground state
is bounded by $\sigma$ (up to an overall factor which plays no role).
There is no need to study the stability of the ground state 
in the transitory regime from underdamped to the overdamped modes
(for $p_{phys} > \llv$), because whatever transitions happened 
is suppressed by a factor $e^{-\int \gamma dt } 
\simeq \exp(-1/\sigma(2n + 1)) \ll 1$. This completes the proof
that scale separation guarantees adiabaticity. 

%
%
%

\subsection{Dissipation in Black Hole metric}
\label{DBHm}

In spite of the fact that the background metric is no longer homogeneous,
the effects of dissipation on Hawking radiation can be studied 
along lines similar to the above.

For simplicity we consider only spherically symmetric and 
stationary BH metrics.
We also choose the unit vector field $l^\mu$ to be stationary and associated with
Freely Falling observers which start at rest at infinity. In this case, 
the expressions for both the metric and $l^\mu$ simplify using the 
$PG$ (Painlev\'e-Gullstrand) 
coordinates, $t,r,\theta, \phi$. One has
\ba 
ds^2 &=& - dt^2 + (dr - v dt)^2 + r^2 d\Omega^2 \, , \nonumber \\
l_\mu &=& (1,  v(r), 0, 0)\, , \nonumber \\
g_{\mu\nu} &=& - l_\mu l_\nu + \perp_{\mu\nu} \, ,\quad
{\rm with} \perp_{\mu\nu}= {\rm diag}(0, 1, r^2, r^2 \sin^2\theta)\, ,
\ea
where $v(r)< 0 $ is the radial velocity of the FF observers, and 
$t$ their proper time, not to be confused with the "Schwarzschild" (Killing)
time.
From the last equation we learn that the 3-surfaces perpendicular
to $l^\mu$ are simply the Euclidean space. Therefore $h= {\rm det}\perp_{ij}
= r^4 \sin^2\theta$ 
is independent of the expansion of $v$, as it is the case 
in cosmology in Lema\^itre coordinates,
see footnote \ref{intfnote}.

However $PG$ coordinates are not comoving along the 
$l^\mu$ field since the shift is given by $\vec v$. 
To determine the space-time dependence 
of the comoving volume element $\sqrt h_c$, 
we introduce the "comoving" coordinate $r_0 = r_0 (t,r)$ defined by
\be
\int_{r_0}^r {dr' \over v(r')} = t \, .
\ee
By definition $r= r(t,r_0)$ gives the trajectory of the FF observer who started from $r_0$ at $t=0$. Thus, for every $(t,r)$, $r_0= r_0 (t,r)$ gives 
the value of radial coordinate at time $t=0$. Using 
\be
dr - v dt = { \partial r \over \partial r_0}
\, dr_0 = { v(r) \over v(r_0)} \, dr_0 \, ,
\ee
the line element in preferred ($l^i = 0$) and comoving
($g_{0i}= 0$) coordinates $(t,r_0)$ reads
\be
ds^2 = - dt^2 +\Big({ v(r) \over v(r_0)}\Big)^2 \, dr_0^2 + r^2 d\Omega^2 \, ,
\ee
where $r \equiv r(t,r_0)$.
Therefore, 
 the evolution of a comoving volume 
centered along the $r_0$-trajectory is given by
\be
h_c^{1/2}(t,r_0) = 
{ v(r) \over v(r_0)} \times r^2 \sin\theta 
= h^{1/2}_{c\, 2D}  \times r^2 \sin\theta \, .
\ee
As in cosmology, it possesses a well-defined time dependence. 
In the present case, it is governed 
by the (shift) function $v(r)$, and the FF trajectories $r= r(t,r_0)$. 

Because of spherical symmetry, our Gaussian action separates into actions
at fixed angular momentum : 
 $S_T = \Sigma_{l,m} \, S_T(l,m)$, see eq. (\ref{psep}). 
Each of them contains 
two-dimensional fields $\phi_{l,m}(t,r)$ and ${\Psi}_{l,m}(t,r)$.
We here consider only the s-wave sector, and drop the $0,0$ index. 
We work with the rescaled field $\tilde \phi = r \phi$, 
$\tilde{\Psi} = 
r 
\, {\Psi}$
and drop the tilde. 
Taking into account the 4D character of the problem 
and eqs. (\ref{covl}, \ref{covl2}), the action 
reads
\ba
S_T &=& {1 \over 2} \int\!\!\!\int dt dr \, r^2  \Big[ \big( 
(\partial_t + v \partial_r)\, {\phi
\over r} \big)^2 - \big( \partial_r \, {\phi
\over r} \big)^2 \Big] \nonumber\\
&& +  {1 \over 2} \int\!dk \int\!\!\!\int
dt dr \, r^2 
\Big[ \big( (\partial_t + { v \partial_r + \partial_r v \over 2  }) {\Psi_k\over r} \big)^2 
-   \big( (k \llv \pi) 
 {\Psi_k\over r}  \big)^2 \Big] 
\nonumber\\
&& + g \llv \int\!\!\!\int 
dt dr \, \, r^2 
\Big[\big(   ({\partial_r \over \llv })^{n+1}\,  {\phi
\over r} 
  \big) \big( 
(\partial_t + { v \partial_r + \partial_r v \over 2  }
) \int\!dk {\Psi_k\over r}\big)\Big]\, .
\ea
By varying this action, there is no difficulty to get the 
equations of motion. One verifies in particular 
 that  the kinetic action of the (rescaled)
 $\Psi^r = h_{c\, 2D}^{1/4} \tilde \Psi
 $ field is as in 2D
 flat space when expressed in preferred coordinates $t,r_0$:
 \be
 S_{\Psi} = {1 \over 2} \int\!dk \int\!\!\!\int
dt dr_0 
\Big[ \big( 
\partial_t{\Psi^r_k}  \big)^2 
-   \big( (k \llv \pi) 
 {\Psi^r_k}  \big)^2 \Big] \, .
 \ee
This means that the Chelva field $\Psi^r_k(t,r)$ is a collection of independent 
oscillators labelled by $r_0$. In other words,
$\Psi^r_k(t,r)$ depends on $t$ and $r$ as 
\be
\Psi^r_k(t,r) = \Psi^r_k(t; r_0(t,r)) \, .
\ee
One also verifies
$
l^\mu \partial_\mu \Psi^r_k(t,r) = \partial_t \Psi^r_k(t; r_0
)\vert_{r_0}\, ,
$
because $l^\mu\partial_\mu r_0(t,r)=0$, by definition.
These facts 
guarantee that eq. (\ref{covcurveddelta}) 
applies.

By a direct analysis of the equation of motion, one obtains two important properties. First, one verifies that there is no "dilution" 
of the dissipation rate as time passes.
This is non trivial (see \cite{Jacobson:1999ay}
  for an {\it a priori} similar system
in which a dilution does occur).
%
%
%
This steady behavior is 
due to the fact that the 
Chelva  fields form a 3 dimensional dense set 
(as opposed to the discrete lattice of \cite{Jacobson:1999ay})
which allows the various $h_c$ to 
cancel each other in the determination of the dissipation rate. 
Moreover one also verifies that stationarity is preserved, not only in the
dissipative aspects, but also in the driven properties govern by 
the kernel $\bf N$, eq. (\ref{n}). 
This is non trivial either since $h_c$ depends on $t$ and appears
in the action $S_{\Psi \phi}$ when working with the rescaled fields $\Psi^r$.
%

With these two results, one can study the impact of dissipation 
on Hawking radiation by applying the technics of \cite{Brout:1995wp}
which were further developed in \cite{Balbinot:2006ua}\cite{Jacobson:2007jx}
to study the impact of dispersion.
When the two scales are well separated, i.e. when 
$\kappa /\llv \ll 1$, 
where $\kappa = \partial_r v$ evaluated at the horizon $v= -1$, 
one can work in the near horizon approximation, wherein  $v = 
-1 + \kappa x$, and in the $p$-representation.  
In this representation, the velocity profile becomes a  
non-trivial operator $\hat v = -1 - i \kappa \partial_p$, 
in fact the only one. When working at fixed Killing frequency $\om_K$,
one obtains 
the radial propagation of the interacting degrees of freedom in terms of an adiabatic evolution of Green functions 
and kernels written in the $p$-representation. These 
are given in Appendix B, with the frequency $\om = i\partial_t$ of that 
Appendix replaced by the FF frequency $\Om = i\partial_t + i v \partial_r
= \om_K - \hat v p$. 
We are planning to report on this soon.

  \bigskip 
    
{\bf Acknowledgements.}    
\noindent
I am grateful to Dani Arteaga and Enric Verdaguer for common work
allowing me to deepen my understanding of dissipative effects. 
I am also grateful to Ted Jacobson for 
countless discussions and remarks. 
I am thankful to many people for having had the opportunity to present this work 
starting with the IAP in October 2004.
This work has been supported by the European Science Foundation network
programme "Quantum Geometry and Quantum Gravity".

 \bigskip 

 \section{Appendix A: Two-point correlation functions}
 
 Since the models we consider are Gaussian, 
  {the} complex function of eq. (\ref{Gf}) 
  governs {\it all} observables 
  built with the Heisenberg field operator $\phi$.
  To analyse it, 
  as  mentioned in the text, 
  it is appropriate to study separately the
  commutator and the anti-commutator. 
  
  We start with the simple part, the commutator
  \be
  G_c(t,t';p)\, \delta^3({\bf p}-{\bf p}') 
  \equiv {\rm Tr}[ \rho_T \, [\phi_{\bf p}(t) , \phi_{\bf p'}^\dagger(t')]_-\, ]\, .
  \label{Gct}
\ee
It possesses several 
properties. 
First, using eq. (\ref{trues}) one sees that 
it decomposes into two terms, one due to the 
non-commuting character of $\phi^d$, the other 
due to that of $\Psi_i^0$. Second, since both 
commutators are c-numbers, it is independent
of $\rho_T$, the state of the system. Hence, for all Gaussian models, one has
\be
G_c(t,t';p) =  [\phi^d(t) , \phi^d(t')]_-\, 
+ \int\!\!\!\!\int\!dt_1 dt_2 \,  G_r(t, t_1) G_r(t',t_2) D(t_1,t_2)\, ,
\label{Gcgs}
\ee
where the ''dissipative'' kernel $D(t_1,t_2)$ is given by
\be
D(t_1,t_2)=
\Sigma_i \Sigma_j \, g_i(t_1) \,  g_j(t_2) \,  [\Psi_i^o(t_1) ,\Psi_j^o(t_2)]_- 
= \Sigma_i \, g_i(t_1) G_{c, i}^o(t_1, t_2) g_i(t_2) \, .
\label{d}
\ee
Notice how this kernel combines the various couplings constants 
and the non-commuting properties of the environment.

The third 
property is the most relevant for us (and perhaps also less often mentioned).
To all orders in $g_i$ and
for all values of $g_i, \Omega_i$ (even with arbitrary
time dependence), 
the following identity 
\be
i \partial_t G_c(t,t';p)\vert_{t=t'} = 1 \, ,
\label{i1}
\ee
holds because it corresponds to the ETC of eq. (\ref{etc}). 
The $1$ on the $rhs$ is guaranteed by the
Hamiltonian character of the evolution of the entire system $\phi + \Psi$.
It is therefore this equation which replaces the constancy of 
the Wronskian 
that was relevant in the case of free evolution. Eq. (\ref{i1}) is crucial for us
because 
the operator $\phi^d(t)$
 exponentially decays in $t-t_{in}$ where
$t_{in}$ is the moment when the interactions are turn on,
since it is an homogeneous solution of eq. (\ref{Gr}). 
Hence the first term in eq. (\ref{Gcgs}) decays
as $exp-\gamma(t+t'-2t_{in})$. Therefore 
at late times with respect to $t_{in}$ in the units of the
inverse decay rate $\gamma^{-1}$,
 the non-commuting properties of field operator
 $\hat \phi$ are {\it entirely
due} to those of the environment degrees of freedom, $\hat \Psi^o$. 
In addition, they sum up {\it exactly} to 1 as shown in the above equation.

We now analyze the anti-commutator, 
\be
G_a(t,t';p)\, \delta^3({\bf p}-{\bf p}') 
  \equiv {\rm Tr}[ \rho_T \, \{\phi_{\bf p}(t) , 
  \phi_{\bf p'}^\dagger(t')\}_+\, ]\, .
\label{Gags0}
\ee
When 
the
density matrix factorizes as $\rho_T = \rho_\phi \,  \rho_\Psi$
before the interactions are turned on
(as it is the case in the "free" vacuum), 
$G_a$ also splits into two terms,
\be
G_a(t,t';p) =  {\rm Tr}[ \rho_\phi \{\phi^d(t) \,, \phi^d(t')\}_+\,] 
+ \int \!\!\!\!\int\! dt_1 dt_2 \, G_r(t, t_1) G_r(t',t_2) N(t_1,t_2)\, .
\label{Gags}
\ee
The first term depends only on the initial state of $\phi$. 
Similarly, 
the driven term depends on the state of the environment
through the ''noise'' kernel
\be
N(t_1,t_2)=
\Sigma_i \Sigma_j  {\rm Tr}[ \rho_\Psi \,
 \{ g_i(t_1)\Psi_i^o(t_1) \,, g_j(t_2)\Psi_j^o(t_2)\}_+\,]  \, .
 \label{n}
\ee
As for the commutator, in the presence of dissipation, 
the first term exponentially decays, expressing the
progressive erasing on the information contained in the initial state
of $\phi$. At late times therefore, as one might have expected,
it is the state of $\Psi$ 
which fixes the anti-commutator of $\phi$. 
This allows to remove the restriction that initially
the density matrices factorizes. If one is interesting by late
time behaviour, only $N$ matters.

In brief, two important results are been obtained.
 First, 
even 
when 
$\phi$ reduces to its
driven term, the second term of eq. (\ref{trues}), 
it still behaves as a canonical degree of freedom in that it exactly
obeys eq. (\ref{i1}).
Second, 
 only two (real) c-number quantities determined by the environment
govern the two-point functions of $\phi$, namely
the kernels $D$ and $N$ of eqs. (\ref{d}) and (\ref{n}). Therefore the set
of 
environements (Gaussian or not)
possessing 
the same kernels will give rise to the same 2pt functions for $\phi$.
Hence they should be viewed as forming an equivalent class. 
The degeneracy can 
be lifted by considering correlations with 
observables containing the operators $\Psi_i$,
or higher order correlations functions of $\phi$ 
(for non-Gaussian environements),
two possibilities we shall not discuss 
in this paper.

 In order to be able to compute 
   $G_c$ and $G_a$, 
  two different routes 
  can be adopted. 
  When $g_i(p)$ and $\Omega(p)$ are constant, 
  one should work in Fourier transform because 
  the equations can be algebraically solved, in full 
  generality. 
  Instead when $g_i(p)$ and/or $\Omega(p)$
  are time-dependent, as it will be the case 
  in expanding universes, 
 it becomes imperative 
 to choose the set of $\Psi_i$'s and
  their frequency $\Omega^2_i(p)$ 
  so as to simplify the time dependence of the equations. 
%
In the body of the article, we proceed with time dependent approach since it 
will allow us to combine general covariance and dissipation.
We present below 
the Fourier analysis which is straightforward and 
well known~\cite{1976AnPhy..98..264A,QNoise}. We encourage
the reader unfamiliar with the treatment of dissipative effects in 
Quantum Mechanics to read it.

\section{Appendix B : \\Stationary states, 
and vacuum 2-point functions}

   In this Appendix, we present the 
  relationships between $G_c$, $G_a$ which always hold
  in stationary states. 
In these cases,
the Green functions $G_c, \, G_a$ 
and the kernels $D, \, N$
are functions of $t-t'$, and
are related to each other in a fundamental way,
generally refered as a Fluctuation-Dissipation relation.
We briefly explain its origin and its physical implications
in the present context. We start the analysis by 
with the most basic object: the retarded Green function $G_r$.

  
  \subsection{The retarded Green function}
  
  The Fourier transform of eq. (\ref{trues})
  gives
  \be
   ( -\om^2 + \om_p^2 ) \phi_{\bf p}(\om) = 
  \Sigma_i  g_i(p) 
  \Psi_i^o({\bf p}, \om)
  + \Sigma_i  g_i^2(p) 
  R_i^o(\om; p)
    \phi_{\bf p}(\om) \, ,
    \ee
   where
   \be
   R_i^o(\om; p)= \left(-(\om + i \e)^2 +\Om_i^2(p))   \right)^{-1} \, ,
    \label{Grfpsi}
    \ee
    is the Fourier transform (defined as in eq. (\ref{FTdef})) 
    of the retarded Green function of $\Psi_i$.
   As usual, its retarded character is enforced by the
   imaginary prescription of the two poles to lay
   in the lower half plane ($\e > 0$).  
   The solution of the above equation is
   \be
    \phi_{\bf p}(\om) =   \phi_{\bf p}^d(\om) + G_r(\om, p)
  \Sigma_i  g_i(p) 
  \Psi_i^o({\bf p}, \om) 
  \label{truef}
     \, , 
    \ee
  where the  Fourier transform of the retarded Green function 
  of $\phi$, the solution of eq. (\ref{Gr}), always takes the form
  \be
   G_r(\om, p)  = \left( -(\om + i \e)^2 + \om_p^2 + \Sigma_r(\om,p)
   \right)^{-1}
   \, .
   \label{Grf}
    \ee
  All effects of the coupling to the $\Psi_i$'s are thus encoded
  in the (retarded) self-energy $\Sigma_r(\om,p)$. For Gaussian theories, 
  it is {\it algebraically} given by
  \be
  \Sigma_r(\om,p)= - \Sigma_i \,  g_i^2(p) R_i^o(\om, p)\, .
  \label{Sr}
  \ee
  The dissipative effects are governed by the imaginary 
  part of $ \Sigma_r(\om,p)$.
  In the present case, one has
  \be 
2{\rm Im} \Sigma_r(\om,p) = - \Sigma_i \, g_i^2(p) G^o_{c, i}(\om, p) 
= - D(\om, p)\, .
\label{DSr}
\ee
To get the first equality we have used the fact that 
in stationary states the retarded Green function and the commutator 
are related by $2$Im$G_r(\om) = G_c(\om)$ for all degrees of freedom, 
free or interacting. In the second equality, we have introduced $D(\om)$,
the Fourier transform of the kernel of eq. (\ref{d}).
 
Several observations should be made here. 
First, from eq. (\ref{Grfpsi}), we obtain that $D(\om)$ is proportional to
$
\Sigma_i g_i^2 \delta(\om - \Omega_i)$. Therefore there is no dissipation
for lower frequencies than the lowest value of $\Omega_i$. 
This simply follows from energy conservation. 
Second, to
obtain "true" dissipation, $D(\om, p)$ should be a continuous
function and not a sum of delta. 
  This can only happen
  when the $\Psi_i$
  form a dense ensemble.
In 
the body of the paper, we shall thus replace the discrete sum on $i$ by an integral
  on a 
  continuous variable, $k$. 
  We shall not consider the discrete cases
even though these could display interesting properties.  
 Third, from a phenomenological point of view, only $D(\om, p)$ matters. 
 Hence we cannot separately know 
 what is the spectrum of the environment, which is given
 by $R_i^o(\om, p)$, and what is the coupling strenght $g_i^2(p)$.
 This is a good thing, because 
 when working in time-dependent settings, we shall exploit this equivalence
 by choosing the simplest model of $\Psi_i$'s which {\it delivers} the 
 kernel $D(\om, p)$ we want.

It is also worth noticing that the dispersive (real) effects
are not directly related to $D$ (or $N$). These 
are governed by the even part of $ \Sigma_r(\om,p)$ which is given by
\be 
 {\rm Re}\, 
\Sigma_r(\om,p) = \int\! \frac{d\om'}{2 \pi}
 \frac{ D(\om')}{\om - \om'}\, ,\label{Kra}
\ee
 where the integral should be understood as a principal value. 
This 
integral relationship
explains why one often founds that 
dispersive effects appear before dissipative effects (for increasing $\om$).
We also learn that the dispersive models studied in the litterature are 
incompatible with the above relations since they assume both 
Re$\Sigma_r \neq 0$ and 
 ${\rm Im} \Sigma_r \equiv 0 $. Therefore these models
 cannot be viewed as resulting from dynamical processes.

  
  \subsection{Fluctuation-Dissipation relations and vacuum self-energy}
  
In this subsection, we derive the
 relationships between $G_c$, $G_a$ and $\Sigma_F$ which
 exist in the true (interacting) ground state.

In interacting theories, the only stationary states are thermal
states, see e.g. \cite{Anglin:1992uq}. 
In these states, 
the Fourier transform
of $D$ and $N$ are related by
\ba
N(\om) &=& D(\om)\, 
\coth(\beta \om/ 2) \, , \nonumber\\
&=& D(\om)\, {\sign(\om)} \, [2n(\vert \om \vert) + 1 ] \,  .
\label{fdur}
\ea
In the second line, $n(\om)$ is the Planck distribution. It gives
the mean occupation number of $\Psi^o_i$ quanta
as a function of the frequency (measured in the rest frame of the bath).
The above relation directly follows from the fact that
the individual commutators and anti-commutators
of the free fields $\Psi^o_i$ 
obey this relation, as any free oscillator 
would do. 
It implies that
the Fourier transform of 
$G_c$ and $G_a$ are also related by
\be
G_a(\om) = G_c(\om)\, {\sign(\om)} \,[2n(\om) + 1 ] \,  .
\label{fd}
\ee
It should be stressed that this equation is exact, 
i.e. non-perturbative, and valid for all theories, 
Gaussian or not. (It indeed directly follows from the cyclic properties of 
the trace defining $G_{\beta}(t,t')=$Tr$ [e^{-\beta H_T} \phi(t) \phi(t')]$). 
 
For Gaussian models, there exists an alternative direct verification
of eq. (\ref{fd}). 
In fact, it suffices to note that 
in steady states the decaying terms of eqs. (\ref{Gcgs})
and (\ref{Gags}) 
play no role, and that 
the Fourier transform of the driven terms are  
respectively given by
\ba
G_c(\om) &=& \vert G_r(\om)\vert^2 \,  D(\om) \, , 
 \label{Gcf}
 \\
G_a(\om) &=& \vert G_r(\om)\vert^2 \, N(\om) \, ,
\label{Gaf}
\ea
since the Fourier transform of the
retarded Green function obeys $G_r(\om)=G_r^*(-\om)$, see eq. (\ref{truef}).
Irrespectively of the 
complexity of $G_r$, i.e. irrespectively of the functions $g_i(p)$, 
$\Omega_i(p)$ and the set of the $\Psi_i$
fields, 
$G_c$ and $G_a$ are thus related to each other
by the FD relation (\ref{fd}).


These universal relations will be relevant 
for to inflationary models wherein only the ground state contributes.  
In particular, they
imply that {\it 
in the true vacuum}, i.e.
when $n(\om)= 0$,
$G_c$ and $G_a$ are exactly related by
$G_a(\om) = G_c(\om)\,\sign(\om) $. Hence the
Wightman function 
\be
 G_W = {1 \over 2} ( G_c + G_a) =  \theta(\om)\,G_c \, ,
 \label{Gwf}
\ee
is determined by the commutator and contains only positive frequency, 
as in the free vacuum. 
Equations (\ref{Gcf},\ \ref{Gaf}) also
allow to compute the 
vacuum self-energy 
of the Feynman Green function in eq. (\ref{GfF}). For
Gaussian models it is  given by
\be
2 {\rm Im}  \Sigma_F(\om) = - D(\om) \, {\sign(\om)} = 
2  {\rm Re}\, \Sigma_i g_i^2 G^o_{F, i}(\om) \, , 
\label{SfD}
\ee
where the $p$ dependence has not been not explicitized, and where we have
taken the real part because of the $i$ in the numerator of eq. (\ref{GfF}).  
With the last equality we recover the fact that in the vacuum, it is sufficient
to consider Feynman Green functions. In non-vacuum states, and in 
non-stationary situations, this is no longer true, thereby justifying
the use of the $in-in$ machinery (also called Schwinger-Keldish formalism),
see e.g. \cite{Arteaga:2007us}.

Before specializing to a specific class of models 
giving rise to dissipation at high frequency, 
we make a pause by asking the following important question:
What should be known 
about the $\Psi_i$ fields to get eqs. (\ref{Gcf}, \ref{Gaf}, \ref{Gwf}) ?
We have proven that it is sufficient for the 
$\Psi_i$'s to be canonical fields, but is it necessary ?

The answer is two fold. 
On one hand, the $\Psi_i$ cannot be stochastically fluctuating
(i.e. commuting) quantities because this would lead
to a violation of eq. (\ref{fdur}) that would imply the
violation of 
eq. (\ref{fd}) and the ETC eq. (\ref{i1}).\footnote{
This constitutes the simplest proof that it is 
inconsistent to couple quantum variables to stochastic (or classical)
ones. If one does so, the ETC of the dressed quantum
variables will always be dissipated after a time of the order
of $\gamma^{-1}$. One can therefore view the experimental
evidences for the ETC of some degrees of freedom 
as a very strong indication that {\it all} 
dynamical variables in our world are quantum mechanical in nature.
This line of thought has been used by W.~Unruh 
to prove that gravitational waves must be quantized.}
They cannot be either a combination of quantum and
stochastic quantities because this would still lead
to a violation of the ETC.
Hence they must be built only from quantum (canonical)
degrees of freedom.

On the other hand, the $\Psi_i$'s can be composite 
operators,\footnote{I am grateful to Albert Roura for 
bringing my attention to this question.} i.e. polynomials of some 
(unknown) canonical fields.
Indeed, 
their commutators would still be all related
to their anti-commutators by the 
FD relation eq. (\ref{fdur}), and this even though  
both depend non-linearly on $n(\om)$ in non-vacuum states.  
The difference with Gaussian models is that 
these non-linear operators possess 
non-vanishing connected higher order correlation
functions. Hence, 
the self-energies $\Sigma_r, \, \Sigma_F$
will contain higher a series in powers of 
$g_i^2$, and not just a single quadratic term as in
eq. (\ref{Sr}, 
\ref{SfD}).
Nevertheless 
these higher loops corrections 
preserve
the validity of eq. (\ref{Gwf}) in the ground state, 
 as well as that of eqs. (\ref{Gcf}, \ref{Gaf}) in any thermal state, when
properly undertood, i.e. with $D$ now defined by -2Im$\Sigma_r$ (as the 
effective dissipation kernel), and $N$
related to it by the FD relation. 


In brief, we have reached/recalled the following 
results.
Firstly, 
the q-number combination ${\Xi}=\Sigma_i g_i\Psi_i$, 
the fluctuating source term of $\phi$, must obey 
the 
FD relation (\ref{fdur}).
This can either be postulated, 
or better, 
be viewed as resulting from the fact that
${\Xi}$ is entirely made out of 
quantum 
degrees of freedom.
Secondly, to lowest order term in $g$, the self-energy
can be obtained by treating ${\Xi}$
as a 
quantum Gaussian variable, whatever its composition may be. 
Thirdly, when dealing with non-Gaussian theories, once 
having computed $\Sigma_r(\om)$, the resulting equations
for the 2-point functions have the same structure and the
same meaning as in Gaussian theories, with $D$ replaced
by -2Im$\Sigma_r$. Therefore, {the entire phenomenology of 
two-point functions can be described 
with Gaussian settings}.

\subsection{The double limit: $g^2 T \to \infty$ followed by $g^2 \to 0$.}

To perform a phenomenological analysis, 
 we need to 
 understand
 how the theory behaves in transitory regime from dissipative 
 to free propagation.
Similarly, 
to study primordial spectra in inflation or Hawking radiation,
we also need to understand how free motion emerges as the 
proper frequency get red-shifted. 
It is therefore relevant 
to study the 
behavior of the two-point function in the 
following double limit. 

One first takes $g^2 T \to \infty$, 
where $T = t-t_{in}$,
$t_{in}$ being 
the moment when the interactions are turn on, and $t$ the moment when 
one studies the field properties. 
This limit implies that the decaying term in 
eq. (\ref{trues}) plays no role. Therefore, near time $t$, 
 the Heisenberg field $\phi(t)$ is a composite operator 
 which only acts in the Hilbert space of $\Xi$.
 
 Secondly, one considers the "free" limit $g^2 \to 0$ of that composite operator. 
 One could naively conclude that $G_c$ and $G_a$ of eqs. (\ref{Gcf},\, \ref{Gaf})  
 would vanish since
 both $D$ and $N$ are proportional to $g^2$. However, this is not the case, 
 because the prefactor  in these equations, $\vert G_r \vert^2$,
 is singular in this second limit. In fact, one verifies that it scales in
 $1/g^2$ in such a way that, in the (interacting) vacuum, one always recovers
 \be
 G_W(\om)_{g^2 \to 0} = 
 {1 \over 2 \om_p}\, 2\pi \delta(\om - \om_p )\, .
 \label{mie}
 \ee
 This 
 is the standard vacuum fluctuations of a free oscillator
 of frequency $\om_p$.
 
 Two important lessons have been reached. 
 First we learned
 is that even though $\phi$ acts 
 {\it only} on the $\Xi$-Hilbert space, 
 when $g^2 \to 0$, it behaves as if it were a free mode 
 possessing its own Hilbert space, with no reference to $\Xi$-dynamics.
 Secondly, in spite of this, 
 the quantum state in the would be Hilbert space
  is 
  still exactly that of $\Xi$.
Therefore, in stationary situations, the only "souvenir" kept by the composite 
operator $\phi^{driven}$
is the equilibrium distribution $n(\om)$ inherited from its parents. 

 Let us now emphasize that the above limit is relevant for non-Gaussian models
 as well. 
 Indeed, in the limit $g^2 \to 0$, there will always be a value of $g^2$
 sufficiently small that the model can be well approximated by a Gaussian
 model. Therefore the 
 behavior of the 2-point functions in the transitory regime 
 from dissipation to free
 propagation can be analyzed {without}
 restriction by studying Gaussian models (at least in the quasi-static limit). 

\section{Appendix C: \\
Dissipative effects above $\llv$. The Phenomenology}

 We now have all the tools to understand 
 models giving rise to dissipation in the vacuum
  above a critical energy scale $\llv$.
%

If one considers only stationary situations (i.e. static metrics 
  and stationary states), and if one adopts
  a 
  phenomenological point of view, one 
  can simply choose the function $D(\om,p)$ entering
 eq. (\ref{DSr})  and  eq. (\ref{Gcf}) {\it as one wishes}. There is indeed
  no restriction on $D(\om,p)$ besides its constitutive properties, namely 
  being odd in $\om$ and giving rise
  to poles in $G_r$ all localized in the lower half $\om$ plane. 
  In this we have reached our first aim, namely
  identify how to generalyze the free settings so as to 
  incorporate some arbitrary dissipative effects.

  We can thus consider the dispersive models which correspond to 
those defined by  
eq. (\ref{disprel}). They 
 are charaterized by a 
  single term giving rise to dissipation above $\llv$. 
  In the vacuum, they are 
fully specified 
 by the 
imaginary part of the (retarded) self-energy 
  \be
 -{\rm Im}\Sigma^{(n)}_r(\om, p) 
    = {\om \over \llv}
    \, p^2 
   \left({p \over \llv}\right)^{2n}
   = 2 {\om} 
   \, \gamma_{n}\, . 
   \label{iDRn2}
   \ee
   In these models, 
   the decay rate (inverse life time) on the mass shell
   is 
   \be
  { 
  \gamma_{n}} =
  {p \over 2} \left({p \over \llv}\right)^{2n+1} \, .
  \label{decayn} 
  \ee
   To verify it, assuming 
  that Re$\Sigma_r = 0$,
  the two poles of $G_r(\om)$ in eq. (\ref{Grf}) 
  are located in
   \be
   \om_\pm(p) = \pm \sqrt{\om_p^2 - \gamma^2} - i \gamma
   \label{omeff}
   \, .
   \ee
   From this, by inverse Fourier transform $G_r(\om)$,  
    one obtains that the
   decay rate 
   is indeed $\gamma$ in the underdamped regime, for $\om_p^2 > \gamma^2$.
  In the overdamped regime, for $\gamma^2 > \om_p^2 $, the decay rates of the 
  two independent solutions of  $G_r^{-1} \phi_d = 0$
  are $\Gamma_\pm = \gamma \pm \sqrt{\gamma^2 - \om_p^2}$.
  
   One thus obtains the following behavior as $p$ grows.  
   For $p \ll \llv$, $\om_\pm  \simeq p$, and one has
    a free propagation which is slightly damped 
    with a life time in the units of the frequency given by  
   $(\llv/\om)^{n+1} \gg 1$. 
   Instead the opposite regime of high momenta $p \gg \llv$, deep in the overdamped
   regime, the two roots $\om_\pm$ are real
   and the notion of propagation (in space-time) is completely absent. 
   In anticipation to what will occur in inflation or in black hole physics,
   we invite the reader 
   to study the migration of the poles of $G_r$ 
   when extrapoling backwards in time a mode, 
   i.e. as $p$ increases. (Remember that the physical momentum of a mode 
   in cosmology is $p(t) = p_o/a(t)$ where $p_o$
   is the norm of the conserved comoving wave vector,
   whereas near a black hole horizon one finds $p(r)= \om /x$ where $x = r-r_S$
   is the proper distance from the horizon measured in a freely falling frame,
   and $\om$ the conserved Killing frequency measured asymptotically.)

   One could of course generalize the above class
   by considering in eq. (\ref{iDRn2})
   polynomials in $p$ dimensionalized by different UV
    scales. However, unless fine tuning, the phenomenology 
    of the transition from the IR dissipation-free sector to the
    dissipative sector will be dominated a single term. One should also 
    consider the possibility that Im$\Sigma$ strictly vanishes 
    below a certain frequency $\Omega_1$. This
    would be the case when the spectrum of the $\Psi$ fields
    possesses such a gap, see the remarks after eq. (\ref{DSr}).
    
 Having the phenomenology of dissipative unitary models under control
 (with dispersive and dissipative related by Kramers relations, see 
  eq. (\ref{Kra}))
 one could confront particle and astro-particle 
    physics data and put lower bounds on $\llv$ for each $n$, in analogy with what 
    was done for (pure) dispersion in \cite{Jacobson:2005bg}. 

\end{document}